\documentclass[twocolumn,preprintnumbers,superscriptaddress,nofootinbib,aps,prd]{revtex4-1}
\pdfoutput=1

\usepackage{graphicx}
\usepackage{subfig}
\usepackage{mathrsfs} 
\usepackage{amsmath}
\usepackage{amssymb}

\newcommand{\N}{N}
\newcommand{\SUNic}{SU(N)_{ic}}
\newcommand{\Lic}{\Lambda_{ic}}
\newcommand{\ra}{\rightarrow}

\begin{document}
\pagestyle{plain}

\preprint{FERMILAB-PUB-11-275-T}

\title{Chiral Quirkonium Decays}

\author{R. Fok}
\affiliation{Department of Physics, University of Oregon, Eugene, OR 97403}

\author{Graham D. Kribs}
\affiliation{Theoretical Physics Department, Fermilab, Batavia, IL 60510}
\affiliation{Department of Physics, University of Oregon, Eugene, OR 97403}

\begin{abstract}

We calculate the two-body decay rates of ``quirkonium'' states formed from
quirks that acquire mass solely through electroweak symmetry breaking.
We consider $\SUNic$ infracolor with two flavors of quirks 
transforming under the electroweak group (but not QCD) 
of the Standard Model.
In one case, the quirks are in a chiral representation of the 
electroweak group, while in the other case, a vector-like representation.
The differences in the dominant decay channels between 
``chiral quirkonia'' versus ``vector-like quirkonia'' are striking.
Several chiral quirkonia states can decay into the unique two-body resonance 
channels $WH$, $ZH$, $t\bar{t}$, $t\bar{b}$/$b\bar{t}$, and $\gamma H$,
which never dominate for vector-like quirkonia. 
Additionally, the channels $W W$, $W Z$, $Z Z$, and $W \gamma$, are shared 
among both chiral and vector-like quirkonia.  Resonances of dileptons
or light quarks (dijets) can dominate for some vector-like 
quirkonia states throughout their mass range, while these modes 
never dominate for chiral quirkonia unless the decays into pairs of 
gauge or Higgs bosons are kinematically forbidden.

\end{abstract}

\maketitle

\section{Introduction}
\label{sec:intro}

Quirks are fermions transforming under the SM gauge group along with
a new strongly-coupled ``infracolor'' group $\SUNic$ \cite{Kang:2008ea}.
(Earlier ideas were also considered in Ref.~\cite{old-quirks}.)
The scale of infracolor confinement, $\Lic$, 
is assumed to be much smaller than the masses of all quirks,
and so the infracolor-strings have an exponentially suppressed 
rate to break. 
Quirks pairs produced in a collider remain in a bound state 
even when produced with large kinetic energies.
This leads to several interesting collider physics and dark matter 
applications \cite{Kang:2008ea,Burdman:2008ek,Cheung:2008ke,Harnik:2008ax,Cai:2008au,Kilic:2009mi,Chang:2009sv,Nussinov:2009hc,Kribs:2009fy,Kilic:2010et,Carloni:2010tw,Martin:2010kk,Harnik:2011mv}.
(Other work on hidden valley models can be found in 
\cite{Strassler:2006im,Han:2007ae,Strassler:2008fv}.)
Certain kinds of quirks have already been searched for 
at the Tevatron by the D0 collaboration \cite{Abazov:2010yb}.

In this paper, we are mainly interested in quirks that acquire 
mass through electroweak symmetry breaking.  This is unlike the 
original model, Ref.~\cite{Kang:2008ea}, 
where quirks acquired ``vector-like'' masses
independently of electroweak symmetry breaking.  
We are motivated in part by the discovery that chiral quirks
bound in quirky baryons can lead to a viable asymmetric dark 
matter candidate \cite{Kribs:2009fy}.  We do not, however, restrict
ourselves to the specific theory or detailed parameter choices 
of \cite{Kribs:2009fy}.  Instead, we consider general $\SUNic$,
and calculate the meson decay rates for both chiral quirks 
as well as vector-like quirks, demonstrating the experimentally
distinguishable signatures.

At this point we should emphasize that only some aspects of quirky
physics can be calculated (or simulated) with standard collider tools.
In general, quirks can be produced in a standard collider physics
process (for us, weak production), but then the $p_T$ of the quirks
must be shed before the quirks settle down into a low-angular-momentum
state.  This ``spin-down'' process is in general
non-perturbative, with the resulting radiation dependent on the 
relative strengths of infracolor and other couplings of the quirks.  
After spin-down and energy loss, the constituent quirks annihilate, 
causing quirky mesons to decay.  It is solely this last step that 
is our interest in this paper.

The annihilation rate of quirky mesons is proportional to the 
lowest non-vanishing radial derivative of the meson wavefunction 
at zero relative quirk displacement.  This is entirely analogous
to positronium and quarkonium \cite{Barger:1987xg}.
For an $S$ state, this is $|\psi(0)|^2$, while for a $P$ state,
$|\psi'(0)|^2$.   At high orbital angular momentum $L$, this 
wavefunction factor is suppressed.  Ref.~\cite{Kang:2008ea} estimated 
the suppression factor in the annihilation probability scaling as 
$(\beta/L)^{L+1}/L$, where $\beta$ is the quirk relative velocity 
and $L>0$.  Therefore, instead of annihilating immediately, the 
quirky bound states are expected to
emit soft radiation to shed their angular momentum.\footnote{The 
radiation may be in the form of soft photons \cite{Harnik:2008ax}
that can be detected as rings in the $\eta-\phi$ plane in colliders.}
As the quirky bound state reaches a low angular momentum 
state $(L \sim 1)$, the constituent quirks ultimately annihilate;
some quirky meson decay rates for certain vector-like quirks 
have been discussed in 
\cite{Kang:2008ea,Burdman:2008ek,Cheung:2008ke,Harnik:2011mv}.

This paper is organized as follows. We will describe our quirk model 
in Sec.~\ref{sec:model}. 
Next, we present a 
qualitative understanding of the parametric dependencies
of the various decay channels in Sec.~\ref{sec:qualitative}.
We present the formalism to calculate the 
decay amplitudes in Appendix~\ref{sec:me}, along with 
the extensive analytical results for all of our quirkonia 
decay rates in Appendices~\ref{sec:chargedquirkoniumdecays}
and \ref{sec:neutralquirkoniumdecays}. 
Much of our results for neutral quirkonia can be obtained
from earlier results on heavy quarkonia \cite{Barger:1987xg},
which we have compared extensively.
Then, we present numerical evaluations of our results, and
discuss their implications, in Sec.~\ref{sec:BR}.
Next, we calculate the decay rates for vector-like quirkonia
in Sec.~\ref{sec:vectorlike}, comparing and contrasting to the
chiral quirkonia decay results.  We summarize and provide a clear
explanation of which modes have ``chiral enhancement'' in 
Sec.~\ref{sec:chiralenhancement}.  We conclude with a discussion,
identifying the major signals that distinguish chiral quirkonia 
from vector-like quirkonia in Sec.~\ref{sec:discussion}.

\section{Model and Setup}
\label{sec:model}

The model we consider is $\SUNic$ with two flavors in the
representations given in Table~\ref{table:quantumnumbers}.
This is the generalization of the model of Ref.~\cite{Kribs:2009fy} 
to $\N$ infracolors.  We assume $\Lic \ll m_Q$, and 
neglect the infracolor confinement contribution to the quirky
meson masses.  The Lagrangian that gives mass to the quirks
is simply
\begin{eqnarray}
{\cal L} &=& \lambda_U Q H u^c + \lambda_D Q H^\dagger d^c \; .
\label{eq:yukawa}
\end{eqnarray}
Despite the abuse of notation ($Q$, $u^c$, $d^c$), we emphasize
that our quirks are color \emph{singlets}. 
After electroweak symmetry breaking, the quirks acquire masses
$M_{U,D} \equiv \lambda_{U,D} v$.  Writing the electroweak doublet 
as $Q = (u, d)$, 
we can write the quirks in terms of four-component Dirac spinors $U,D$ 
\begin{eqnarray}
U \;=\; \left( \begin{array}{c} u \\ {u^c}^\dagger \end{array} \right) 
& &
D \;=\; \left( \begin{array}{c} d \\ {d^c}^\dagger \end{array} \right) 
\end{eqnarray}
where $U,D$ have electric charge $q = \pm 1/2$.
The quirky mesons formed from these objects include
\begin{eqnarray}
(U \bar{U}), (D \bar{D}) &\quad& \mbox{neutral mesons} 
                                 \label{eq:neutralmeson} \\
(U \bar{D}), (D \bar{U}) &\quad& q = \pm 1 \; \mbox{charged mesons} \; .
                                              \label{eq:chargedmeson}
\end{eqnarray}

There are two interesting regions of parameter space satisfying
the requirement $\Lic \ll M_{U,D}$.  
One occurs when one quirk mass is much heavier than the other, 
$M_U \gg M_D$ or $M_D \gg M_U$, such that 
there is one set of heavy neutral mesons, one set of intermediate-mass 
electrically charged mesons, and one of set of light mesons.  
In this regime, the heavier mesons generically weak decay to the
lightest mesons (microscopically the heavier quirks are weak decaying 
into the lighter quirks) \emph{before} the quirks themselves have time 
to annihilate.  In this regime, the relevant annihilation channels consist 
solely of the lightest neutral mesons.  

\begin{table}[t]
\begin{tabular}{c|ccc}
      & $\SUNic$        & $SU(2)_L$    & $U(1)_Y$ \\ \hline
$Q$   & $\mathbf{N}$        & $\mathbf{2}$ &    0     \\
$u^c$ & $\bar{\mathbf{N}}$  & $\mathbf{1}$ &   $-1/2$ \\
$d^c$ & $\bar{\mathbf{N}}$  & $\mathbf{1}$ &   $+1/2$ \\
\end{tabular}
\caption{Quirk quantum numbers.}
\label{table:quantumnumbers}
\end{table}

The second regime, and the main focus of this paper, 
is when $M_U \simeq M_D$.  When the two flavors of quirks
are very nearly degenerate in mass, all of the mesons given in
Eqs.~(\ref{eq:neutralmeson}),(\ref{eq:chargedmeson}) are 
virtually stable against weak decay.  All of the quirk pairs within 
the mesons therefore annihilate well before the kinematically-suppressed 
weak decay occurs.  This leads to four distinct ``towers'' of mesons:
two sets of neutral mesons and two sets of (oppositely) charged mesons.

The neutral mesons $(U \bar{U})$ and $(D \bar{D})$ can mix with 
each other through infragluon box diagrams that are superficially
similar to the $W$-box diagrams within the SM that lead to mixing among 
the neutral mesons of QCD\@.  However, unlike QCD, all of the quirks are 
heavy, while the gauge bosons being exchanged in the box diagram 
are massless.  This small mixing is an interesting effect for further study.
Our meson decay rates are invariant under $U \leftrightarrow D$,
and we simply compute $(Q \bar{Q})$ as if it were an exact
$(U \bar{U})$ or $(D \bar{D})$ eigenstate.  In practice, 
there may be either a small admixture between these states,
in which case the mixing angle cancels out in our branching ratio 
calculations, or otherwise for maximal mixing, we treat 
$(Q \bar{Q})$ as the $[(U \bar{U}) + (D \bar{D})]/\sqrt{2}$ eigenstate.  

The quirkonium bound state confining potential in the Coulombic 
approximation is \cite{Kribs:2009fy}
\begin{equation}
V(r) \;=\; - \frac{\bar{\alpha}}{r} \; , \label{eq:potential}
\end{equation}
where $\bar{\alpha}$ contains the relevant couplings for the
quirks in our model.  When infracolor dominates, this is
given by 
$\bar{\alpha} \simeq \bar{\alpha}_{ic} \equiv C_2(\mathbf{N}) \alpha_{ic} 
= (\N^2 - 1)/(2 \N) \alpha_{ic}$.
The decay widths are 
proportional to the meson wavefunction when the two constituent 
quirks overlap. The wavefunction factors that appear in the 
decay widths, for $S$ and $P$ states are 
\begin{eqnarray}
|R_S(0)|^2  &=& 4 \left( \frac{1}{4} \bar{\alpha}_{ic} M \right)^3 
                \label{eq:Swave} \\
|R_P'(0)|^2 &=& \frac{1}{24} 
                \left( \frac{1}{4} \bar{\alpha}_{ic} M \right)^5 \; ,
                \label{eq:Pwave}
\end{eqnarray}
where $M$ is the mass of the meson of the appropriate quirkonia state.  

Implicit in evaluating the wavefunctions, 
Eqs.~(\ref{eq:Swave}),(\ref{eq:Pwave}), 
we have assumed the binding energy is dominated by the contributions 
from the infracolor interaction.  This is \emph{not} assumed by
our analytic results, which are written in terms of the
radial wavefunction at the origin.  Moreover, since our 
numerical results are concerned with \emph{ratios} of decay rates, 
the dependence on the wavefunction completely drops out of the 
$S$ state quirkonia decay rates, and is not particularly sensitive 
for $P$ states, as we will see.  

Transitions between principal quantum numbers can occur, 
just as in bound state problems of QED\@.  The transition between
the $n=2$ $P$ states to the $n=1$ $S$ states is given
by the Lyman-alpha electromagnetic transition rate.   
This was estimated for neutral quirkonia to be \cite{Kribs:2009fy} 
\begin{equation}
\Gamma_{{\rm L-}\alpha} = \frac{4}{9} e_Q^2 \alpha_{em} E_{L\alpha}^3 
                    |\langle 0|r|1\rangle |^2 
                  = \frac{1}{4} \bigg( \frac{8}{81} \bigg)^2 \alpha_{em} 
                                       \bar{\alpha}_{ic}^4 M,
\label{eq:Lalpha}
\end{equation}
where $M$ is roughly the meson mass.  Charged quirkonia have the 
same rate, so long as infracolor dominates the potential, 
Eq.~(\ref{eq:potential}).  As we will see, the Lyman-alpha
transition is typically faster than the annihilation rates of
$P$ state quirkonia.  There are exceptions, however, for
chiral quirkonia, which we calculate below, and discuss the resulting 
final state signatures.

\section{Qualitative Description of Results}
\label{sec:qualitative}

The formalism we used for our quirkonia decay calculations
is given in Appendix~\ref{sec:me}, and the complete
analytic results are given in 
Appendices~\ref{sec:chargedquirkoniumdecays} and 
\ref{sec:neutralquirkoniumdecays}.
Here, we evaluate the parametric scaling 
of the various transition and decay rates.  
There are five qualitatively distinct rates 
involved in quirkonia decay:
\begin{eqnarray}
\Gamma_{{\rm L-}\alpha}(P \ra S) 
                      &\sim& \alpha_{em} \bar{\alpha}_{ic}^4 M \\
\Gamma(S \ra g'g')    &\sim& \bar{\alpha}_{ic}^5 M \\
\Gamma(S \ra {\rm SM} + {\rm SM})
                      &\sim& \alpha_{\rm SM}^2 \bar{\alpha}_{ic}^3 M 
                             \left( \frac{M^2}{m^2} \right)^\beta \\
\Gamma(P \ra g'g')    &\sim& \bar{\alpha}_{ic}^7 M \\
\Gamma(P \ra {\rm SM} + {\rm SM})
                      &\sim& \alpha_{\rm SM}^2 \bar{\alpha}_{ic}^5 M 
                             \left( \frac{M^2}{m^2} \right)^\beta 
\end{eqnarray}
The first rate corresponds to the Lyman-alpha electromagnetic 
transition given by Eq.~(\ref{eq:Lalpha}) above.
The second rate, $S \ra g'g'$ refers to specifically the $^1S_0$ 
state decaying into a pair of infragluons.
Note that the $^3S_1$ quirkonia state does not decay into a 
pair of massless gauge bosons, due to angular momentum conservation, 
just like the $^3S_1$ quarkonia state \cite{Barger:1987xg}. 
The third rate, $S \ra {\rm SM} + {\rm SM}$ refers to 
$^1S_0$ or $^3S_1$ state decaying into a pair of SM particles 
with SM coupling $\alpha_{\rm SM}$.
The fourth and fifth rates, 
$P \ra g'g'$ and $P \ra {\rm SM} + {\rm SM}$, refer to any of 
the $P$ states decaying into the above modes.  
Generally, if the Lyman-alpha transition is possible
(i.e., for any of the $P$ states), it dominates over   
the quirkonia decay modes.  The exception to this is if there
is longitudinal or Yukawa ``chiral enhancement'', which can 
occur either singly ($\beta = 1$) or doubly ($\beta=2$) 
depending on the final state.
Double-longitudinal enhancement tends to overcome Lyman-alpha 
emission when $M$ is somewhat larger than one of the SM bosons, 
$m = M_W,M_Z$, compensating for the suppression by the 
larger number of couplings.  
Calculating exactly which modes are enhanced, and why, 
is the main thrust of the paper.

The residual dependence of the decay branching ratios on
the parameters of the theory, $\alpha_{ic}$, $\Lic$, and
the infracolor group $\SUNic$, arise from:
a) neutral quirkonium decay into infraglueballs,
b) possible Lyman-alpha infraglueball transition, and 
c) Lyman-alpha electromagnetic transition.

The first issue, neutral quirkonium decay into infraglueballs, 
is handled by choosing to evaluate ``width ratios'' into visible
SM particles, rather than the standard branching ratios of 
neutral quirkonium decays.  This is because the infraglueballs
are expected to be very long-lived and escape the 
detector \cite{Kang:2008ea},
and thus annihilation to $g'g'$ is expected to yield no hard 
SM resonance signal.  The possibility of 3-body decays,
going into a pair of infraglueballs as well as a SM particle,
is beyond the scope of this paper.  

The second issue, Lyman-alpha infraglueball transition,
can occur if the kinematics of the transition permit it.
The energy difference between the $n=2$ $P$ state and the
$n=1$ $S$ state is $(3/32) \bar{\alpha}_{ic}^2 M$.
The infraglueball mass is of order, but somewhat larger
than $\Lic$.  For $\bar{\alpha}_{ic} \sim O(0.1)$ and 
$\Lic \sim O(1 \; {\rm GeV})$, 
the infraglueball mass is already close to this energy splitting, 
and so kinematic suppression is generic for somewhat larger
values of $\bar{\alpha}_{ic}$.
For smaller values of $\Lic$, presumably accompanied by smaller 
values of $\bar{\alpha}_{ic}$, the infraglueball emission is 
less kinematically suppressed but has an overall transition 
rate that is smaller.  In this region of small $\bar{\alpha}_{ic}$, 
the $P$ states are more likely to transition to $S$ states 
before annihilating. 
A different choice of the infracolor group introduces an order one 
change to $\bar{\alpha}_{ic}$, as well as the running of 
$\bar{\alpha}_{ic}$ in the Coulombic potential \cite{Kribs:2009fy}. 
Since these changes appear as shifts in the coupling of the
Coulombic potential, they can be at least partially absorbed
by a redefinition of $\bar{\alpha}_{ic}$.

Finally, there is the different dependence on $\bar{\alpha}_{ic}$ 
(and $\alpha_{\rm SM}$) between the Lyman-alpha transition rate 
versus the quirkonia decay rates.  The Lyman-alpha transition 
scales as $\alpha_{\rm SM} \bar{\alpha}_{ic}^4$, whereas the decay rates of 
the $P$ states scale as $\alpha_{\rm SM}^2 \bar{\alpha}_{ic}^5$ 
through $|R_P'(0)|^2$.  Hence, there is a relative suppression 
of branching and width ratios of roughly 
$\alpha_{\rm SM} \bar{\alpha}_{ic}$ relative to 
the Lyman-alpha electromagnetic transition rate. 

Consider two regimes, one in which the Lyman-alpha transition rate 
is dominant and the other where it is subdominant to decay rates
that are doubly-enhanced.  In the latter regime, the branching and 
width ratios are rather insensitive to $\bar{\alpha}_{ic}$, since
the only dependence on $\bar{\alpha}_{ic}$ enters from the
wavefunction.  This dependence drops out, analogous to the $S$ states.
In the first regime, however, the total width is dominated by the 
Lyman-alpha transition rate.  The width and branching ratios of 
decay processes scale linearly with $\bar{\alpha}_{ic}$.
But, the regime where the potential is dominated by infracolor 
while still allowing a Coulombic approximation to the 
bound state potential only allows for about one order of magnitude
change in $\bar{\alpha}_{ic}$.

\section{Branching Ratios and Width Ratios}
\label{sec:BR}

We now present our results for charged and neutral quirkonia
decay rates.  
In what follows, we carry out several numerical computations
of branching ratios and width ratios, using the results from
Appendices~\ref{sec:chargedquirkoniumdecays} and 
\ref{sec:neutralquirkoniumdecays}, 
to demonstrate the dominant
SM decay channels for the various bound states of quirkonia.

In the following, we have chosen a specific infracolor group, 
$\N = 2$, and infracolor coupling, $\bar{\alpha}_{ic} = 0.2$.
As we discussed above, the additional relative suppression 
factor of the $P$ state decay rates is only one power of 
$\bar{\alpha}_{ic}$.  The value $\bar{\alpha}_{ic} = 0.2$
tends to maximize the possibility that $P$ states can, 
at larger quirkonium masses, annihilate into SM modes before 
the Lyman-alpha transition occurs.

\subsection{Charged quirkonia}
\label{subsec:charged}

We now discuss the branching ratios of charged quirkonia. 
The analytic results for the decay rates can be found in 
Appendix~\ref{sec:chargedquirkoniumdecays}.  Given a final state $f$, 
the branching ratio for $f$, $BR(Q\bar{Q} \rightarrow f)$, is

\begin{equation}
BR(Q\bar{Q} \rightarrow f) = \frac{\Gamma(Q\bar{Q} \rightarrow f)}{
                               \sum_f \Gamma(Q\bar{Q} \rightarrow f)},
\end{equation}
where the sum is over all final states. 

\begin{figure*}
	\centering
	\subfloat[$^1S_0$]{
		\label{BR1S0c}
		\includegraphics[width=0.48\linewidth]{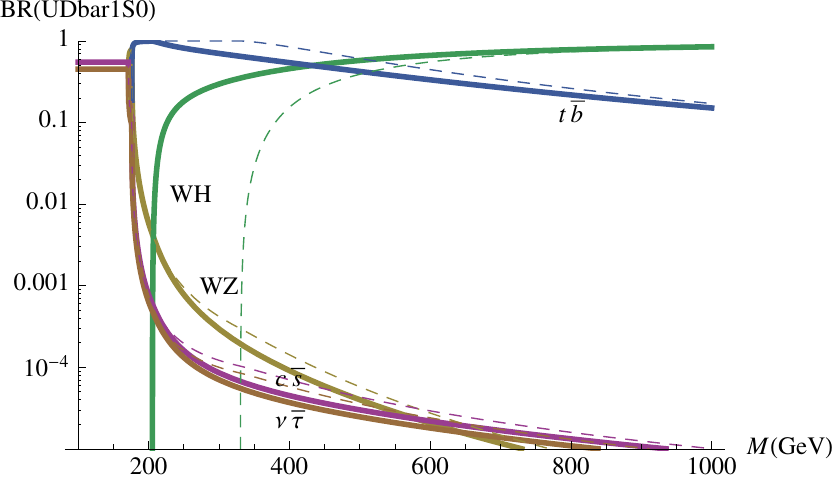}
		}
	\subfloat[$^3S_1$]{
		\label{BR3S1c}
		\includegraphics[width=0.48\linewidth]{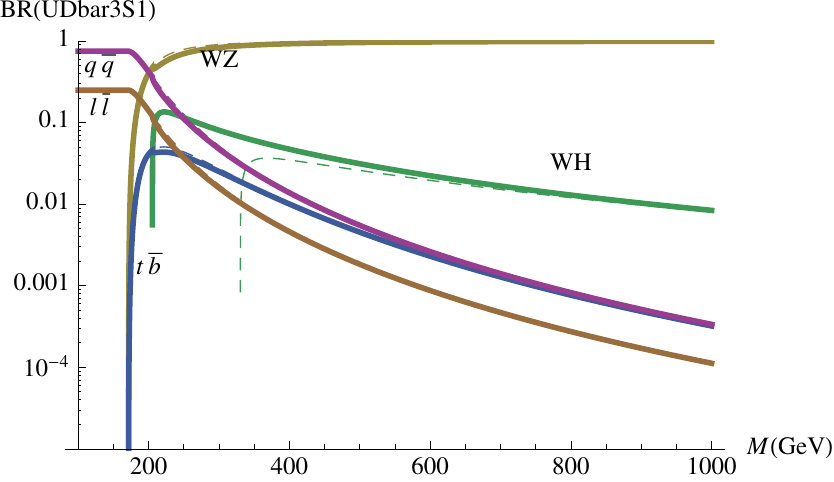}
		}	\\
	\subfloat[$^1P_1$]{
		\label{BR1P1c}
		\includegraphics[width=0.48\linewidth]{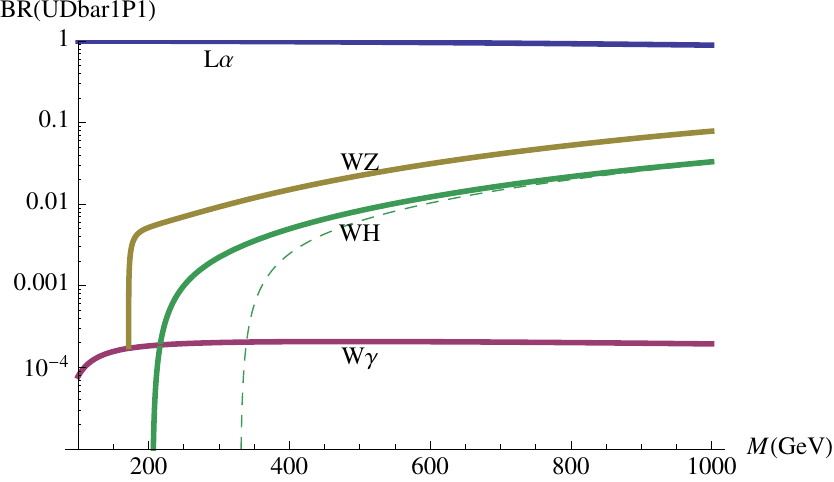}
		}
	\subfloat[$^3P_0$]{
		\label{BR3P0c}
		\includegraphics[width=0.48\linewidth]{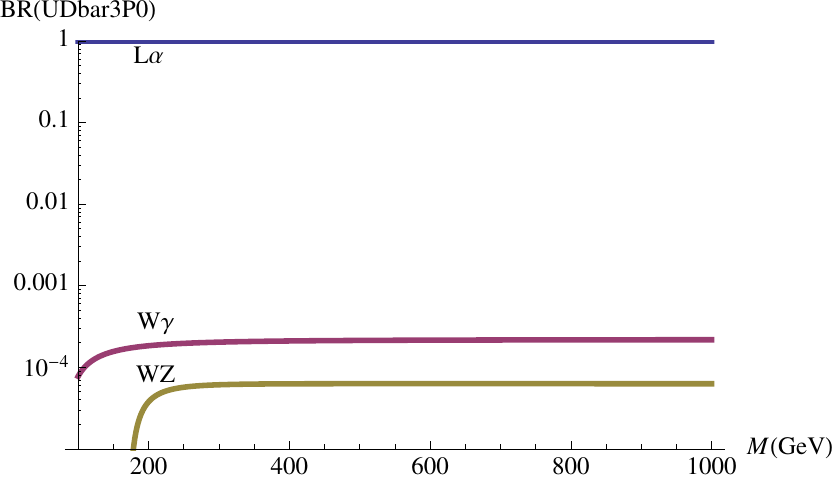}
		}	\\
	\subfloat[$^3P_1$]{
		\label{BR3P1c}
		\includegraphics[width=0.48\linewidth]{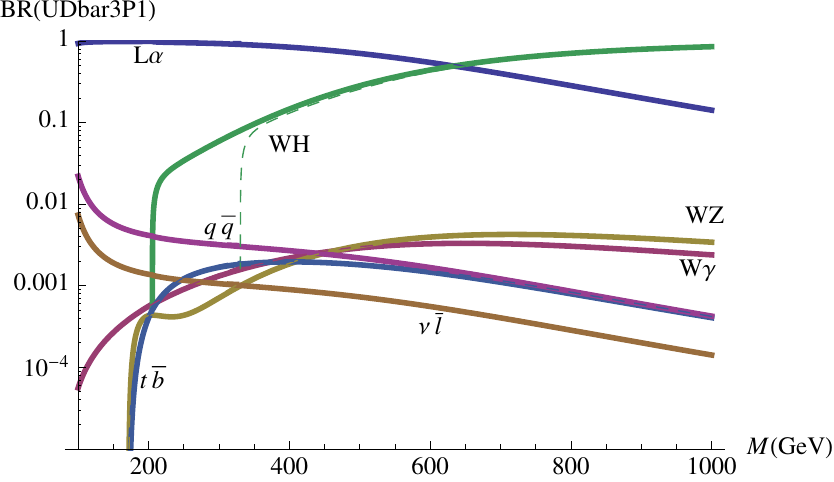}
		}	
	\subfloat[$^3P_2$]{
		\label{BR3P2c}
		\includegraphics[width=0.48\linewidth]{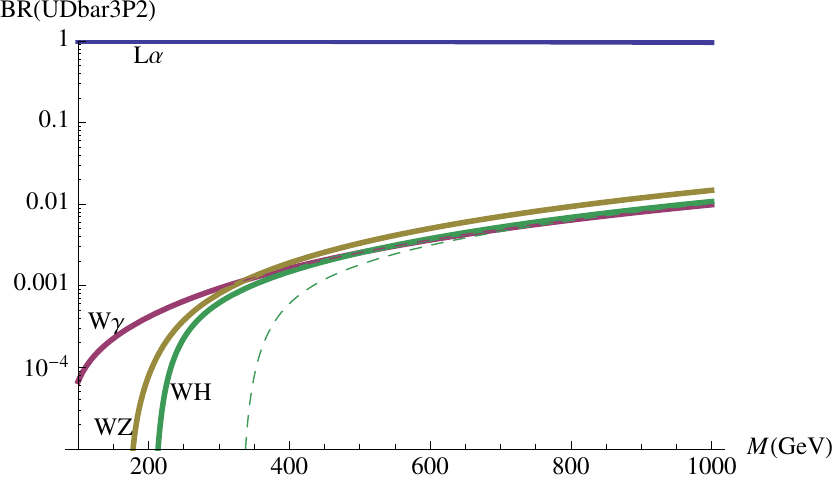}
		}
\caption{Decay branching ratios of charged chiral quirkonia 
in different $J^{PC}$ states. 
Solid lines are with Higgs mass $M_H = 125$ GeV, dashed lines 
with $M_H = 250$ GeV.}
\label{BRc}
\end{figure*}

The charged quirkonium case is particularly simple. As the system 
is electrically charged, it cannot decay into $g'g'$. 
Fig.~\ref{BRc} shows the decay branching ratios of charged quirkonium states. 
For all states, only the $WH$ partial width is sensitive to 
different values of the Higgs mass. The plots shown for charged quirkonia 
here are also applicable to any massive bound states that 
only decay via the electroweak $SU(2) \times U(1)$ group, 
with electric charges $Q_u = -Q_d = 1/2$. Note that we only show 
the summed width over the massless fermions (2 quark pairs, 
3 lepton-neutrino pairs), as the widths of all massless fermions 
are the same (see App.~\ref{ffbar}). Also, we only show the decays 
for the $U \Bar{D}$ meson. We have checked that the widths for 
$\Bar{U} D$ decay are the same.

The branching ratios of different bound states are plotted in 
Fig.~\ref{BRc}. For the $S$ states, the $WZ$ partial width dominates. 
For $P$ states, radiative transition usually dominates. 
For $^3P_1$, the $WH$ width becomes larger than the 
Lyman-alpha transition width when the meson mass is larger 
than $ \gtrsim 600$ GeV, provided that the meson is heavier 
than the threshold. 

\subsection{Neutral Quirkonia}

The results for neutral quirkonia are more complicated than their 
charged counterparts. Not only that there are more decay channels, 
but in some cases the mesons can decay into infraglue pairs, $g'g'$, 
that hadronize into an infraglueball pair $\phi'\phi'$. 

\subsubsection{$g'g'$}
\label{subsec:glueballs}

As the mesons are color singlets, only $t$- and $u$-channels contribute 
to the decay amplitudes of $B \to g'g'$. Then the decay rate should be 
proportional to that of $B \to \gamma \gamma$. A simple calculation 
shows that, for an $\SUNic$ color gauge group,
\begin{equation}
\Gamma(B \to g'g') = \frac{N^2-1}{4N^2} 
                     \frac{\alpha_{ic}^2}{e_Q^4 \alpha_{\rm em}^2} 
                     \Gamma(B \to \gamma\gamma) 
\end{equation}
where $e_Q$ is the quirk electric charge. Setting $N=3$ reproduces 
the results in \cite{Barger:1987xg}. The $^3S_1$ state cannot decay 
into $g'g'$. Its decay into $g'g'g'$ is given by \cite{Cheung:2008ke},
\begin{equation}
\Gamma(^3S_1\to g'g'g') = \frac{(N^2-1)(N^2-4)}{N^2} 
                          \frac{(\pi^2 - 9) \alpha_{ic}^3}{9 \pi M^2} 
                          |R_S(0)|^2 \; .
\end{equation}
This vanishes for the special case $SU(2)_{ic}$, since three gluons cannot 
form a infracolor singlet.

Instead we present our results in terms of a ``width ratio''
\begin{equation}
WR(Q\bar{Q} \rightarrow f) = \frac{\Gamma(Q\bar{Q} \rightarrow f)}{
    \sum_{f \neq \phi'\phi'}\Gamma(Q\bar{Q} \rightarrow f)},
\end{equation}
Also, for reasons of clarity, we do not present the plots for 
the branching ratios when the Higgs mass deviates from 125 GeV\@. 
Unless the final states involve Higgs bosons, a larger Higgs boson mass 
would only push the corresponding thresholds towards higher meson masses, 
leaving the other width ratios mostly unchanged as 
in the case of charged quirkonia. However, there is a qualitative change 
in the width ratios for the $^3P_0$ state when the Higgs mass is 
sufficiently large, which will be discussed below.

\subsubsection{$^1S_0$ and $^3S_1$}

\begin{figure*}
	\centering
	\subfloat[$^1S_0$]{
		\label{BR1S0n}
		\includegraphics[width=0.48\linewidth]{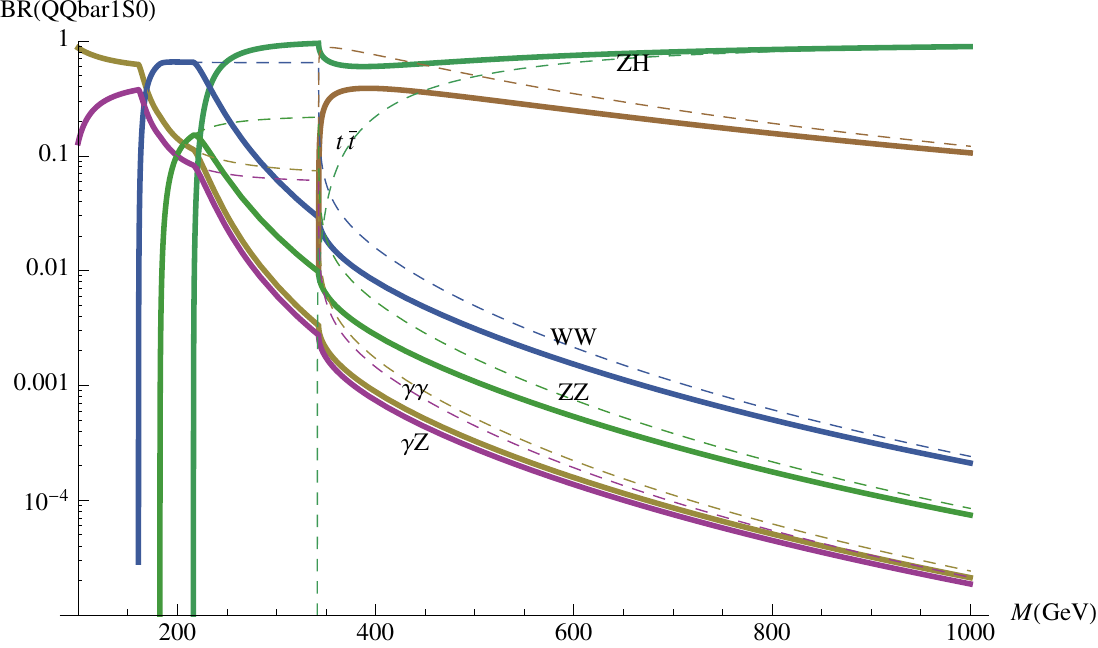}
		} 
	\subfloat[$^3S_1$]{
		\label{BR3S1n}
		\includegraphics[width=0.48\linewidth]{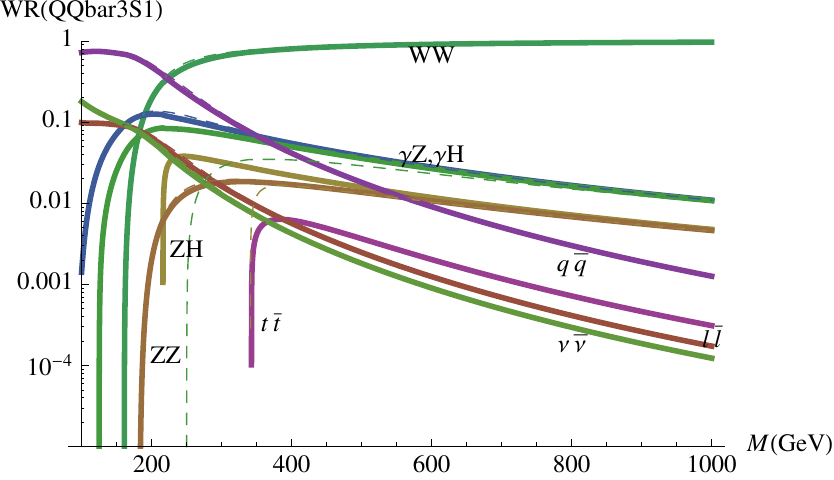}
		} \\
	\subfloat[$^3P_0, M_H=125 $ GeV]{
		\label{BR3P0nogg}
		\includegraphics[width=0.48\linewidth]{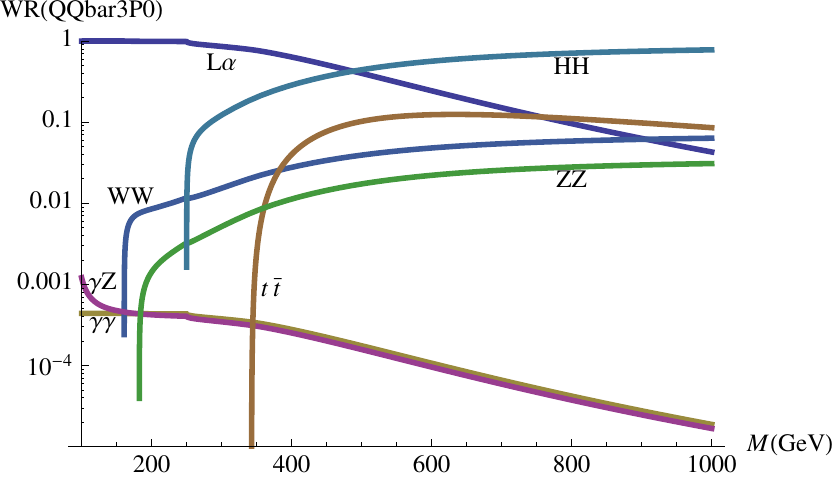}
		}
	\subfloat[$^1P_1$]{
		\label{BR1P1n}
		\includegraphics[width=0.48\linewidth]{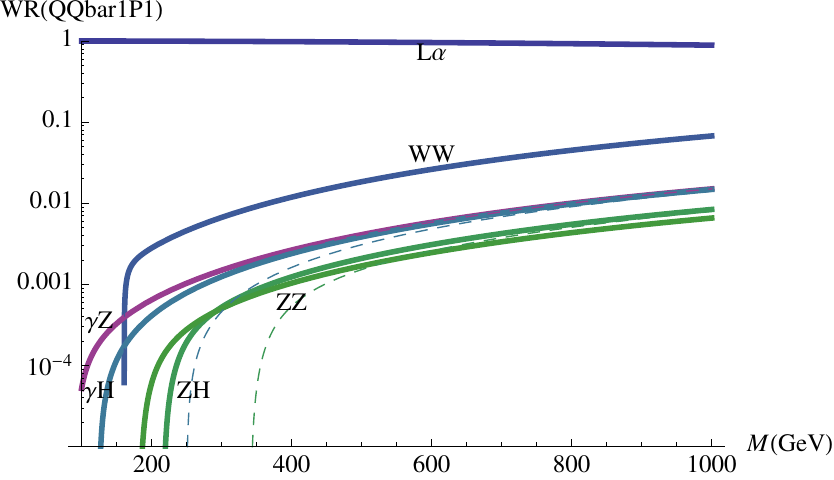}
		} \\ 
	\subfloat[$^3P_1$]{
		\label{BR3P1n}
		\includegraphics[width=0.48\linewidth]{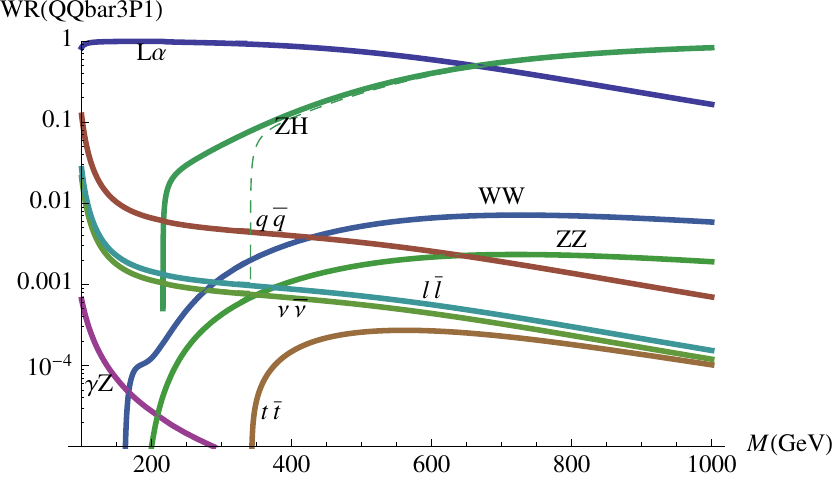}
		} 
	\subfloat[$^3P_2$]{
		\label{BR3P2n}
		\includegraphics[width=0.48\linewidth]{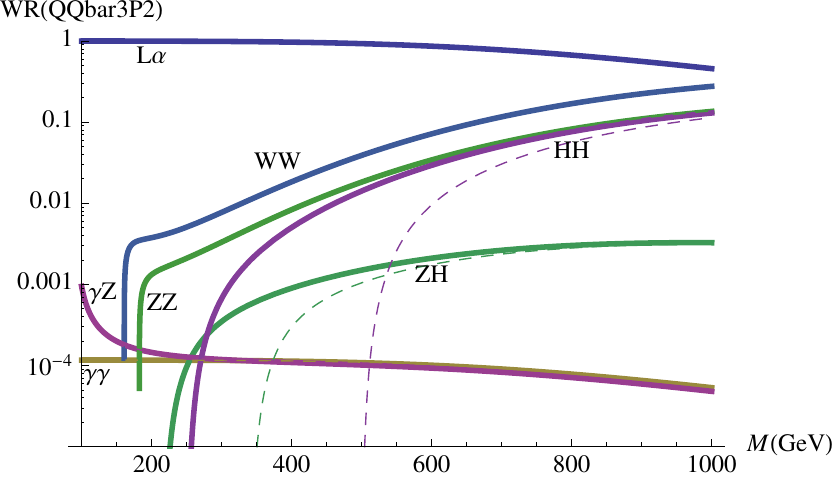}
		} \\

\caption{Decay width ratios of neutral chiral quirkonia in 
different $J^{PC}$ states. 
Solid lines correspond to a Higgs mass $M_H = 125$ GeV, 
while dashed lines correspond to $M_H = 250$ GeV\@.  
In many instances, there is no difference between the width ratios 
for different Higgs masses, and thus the solid lines overlap 
the invisible dashed lines.  For figure (c), we have presented the 
choices $M_H = 125$ GeV\@.  We illustrate the difference in decay width
ratios changing to $M_H = 250$ GeV in Fig.~\ref{BR3P0Higgs250}.}
\label{BRS}
\end{figure*}

The width ratios for the S states are shown in Figs.~\ref{BR1S0n}-\ref{BR3S1n}. 
The dominant decay channels are $ZH$ and $t\bar{t}$. 
The $ZH$ channel receives double enhancement from the 
longitudinal $Z$ and the Yukawa coupling between the Higgs boson 
and the quirk. Also, the $t\bar{t}$ channel is enhanced by the top mass.

The results for the $^3S_1$ state can be discussed more precisely 
because the infraglue channel is absent. Even for values of $M$ not 
far away from $2m_W$, the double longitudinal $WW$ mode dominates. 
Because of Bose symmetry, the two $Z$'s cannot be longitudinal 
simultaneously and the $ZZ$ mode is suppressed. 

\begin{figure}
\centering
\includegraphics[width=0.98\linewidth]{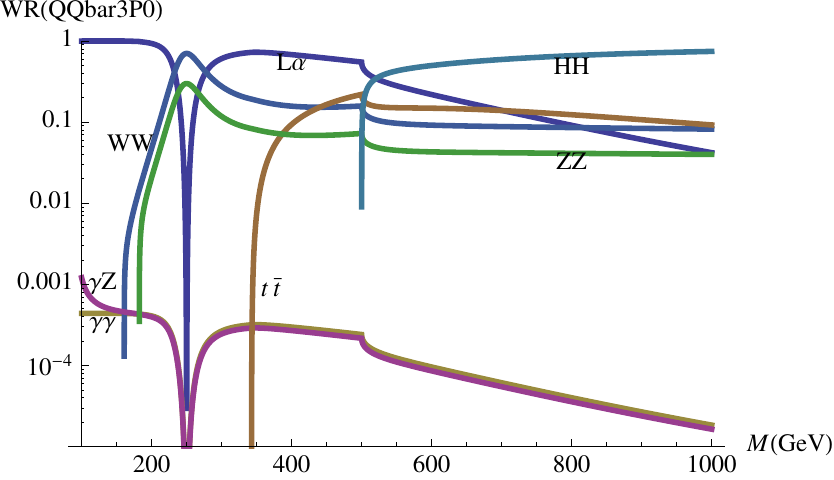}
\caption{$^3P_0$ with $M_H = 250$ GeV. The resonance structure is due 
to the $s$-channel Higgs boson.}
\label{BR3P0Higgs250}
\end{figure}

\subsubsection{$^1P_1$}

The width ratios are shown in Fig.~\ref{BR1P1n} is dominated by the 
Lyman-alpha transition throughout the sub-TeV range. All other widths 
contain a single enhancement factor, from either the longitudinal mode 
or the quirky Yukawa.

\subsubsection{$^3P_0$}

The $^3P_0$ width ratios exhibit an interesting feature when the 
Higgs mass is larger than $2m_W, 2m_Z$ and $2M_t$, where $M_t$ 
is the top mass. The decay channels $WW$, $ZZ$, and $t\bar{t}$ 
involves an $s$-channel Higgs boson exchange. When the meson mass 
is near the Higgs mass $M \sim M_H$, the widths are enhanced 
by the $s$-channel Higgs resonance. This can be seen in 
Fig.~\ref{BR3P0Higgs250}. There, the $WW$ and $ZZ$ widths have 
a resonance at $M = M_H = 250$ GeV when the $s$-channel Higgs boson 
is on-shell.  The $t\bar{t}$ width does not exhibit this behavior 
because at $250$~GeV, the decay into two top quarks from a 
single Higgs boson is kinematically forbidden. 

\subsubsection{$^3P_1$}

The branching ratios for the $^3P_1$ state are shown in 
Fig.~\ref{BR3P1n}.  The $ZH$ channel are doubly enhanced 
and is dominant for $M \gtrsim 700$ GeV\@.

\subsubsection{$^3P_2$}

The channels $WW$, $ZZ$ and $HH$ are doubly enhanced and will 
take over the radiative transition at high meson mass ($\gtrsim 1$ TeV).

\begin{figure*}
	\centering
	\subfloat[$^1S_0$]{
		\label{BR1S0nvector}
		\includegraphics[width=0.48\linewidth]{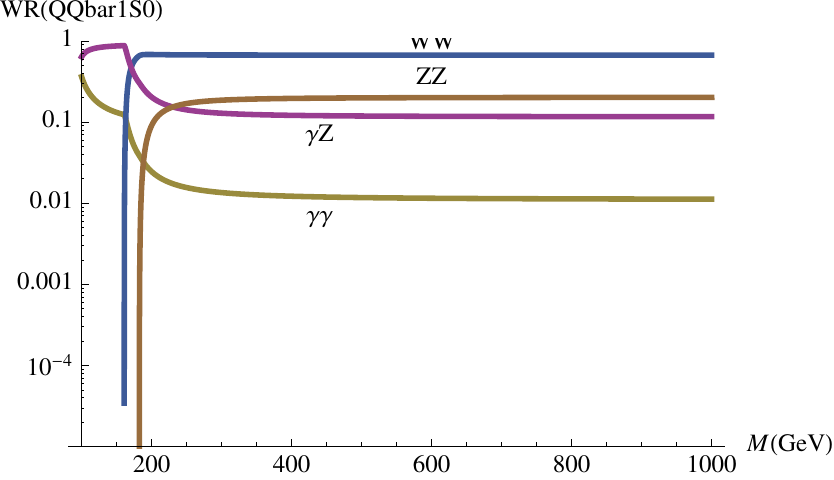}
		} 
	\subfloat[$^3S_1$]{
		\label{BR3S1nvector}
		\includegraphics[width=0.48\linewidth]{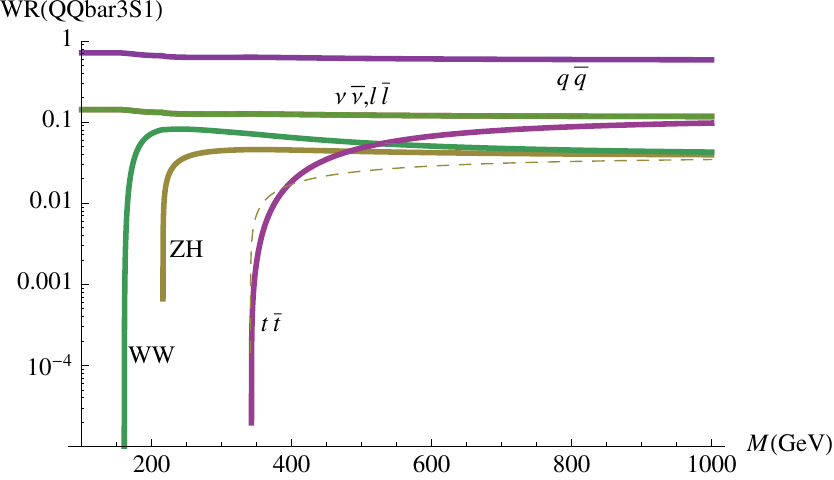}
		} \\
	\subfloat[$^3P_0, M_H=125 $ GeV]{
		\label{BR3P0nvector}
		\includegraphics[width=0.48\linewidth]{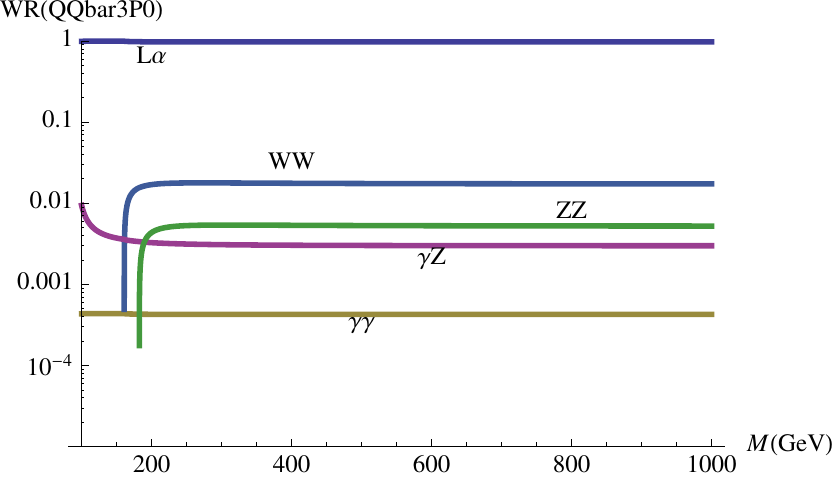}
		}
	\subfloat[$^1P_1$]{
		\label{BR1P1nvector}
		\includegraphics[width=0.48\linewidth]{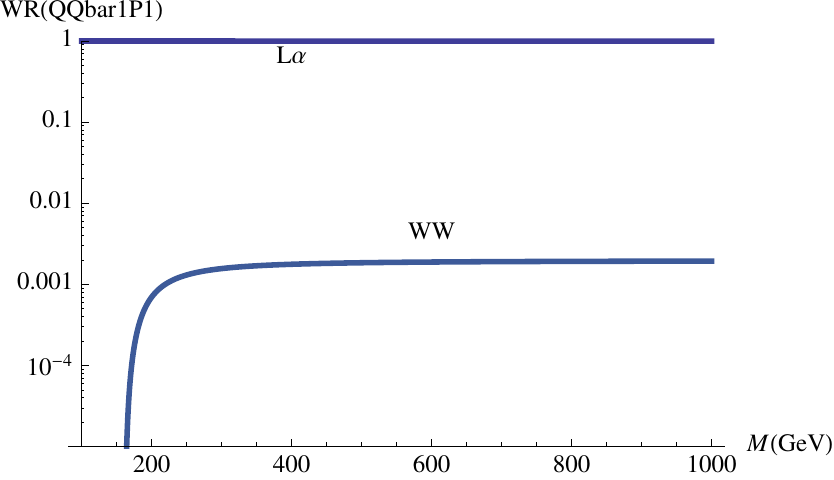}
		} \\
	\subfloat[$^3P_1$]{
		\label{BR3P1nvector}
		\includegraphics[width=0.48\linewidth]{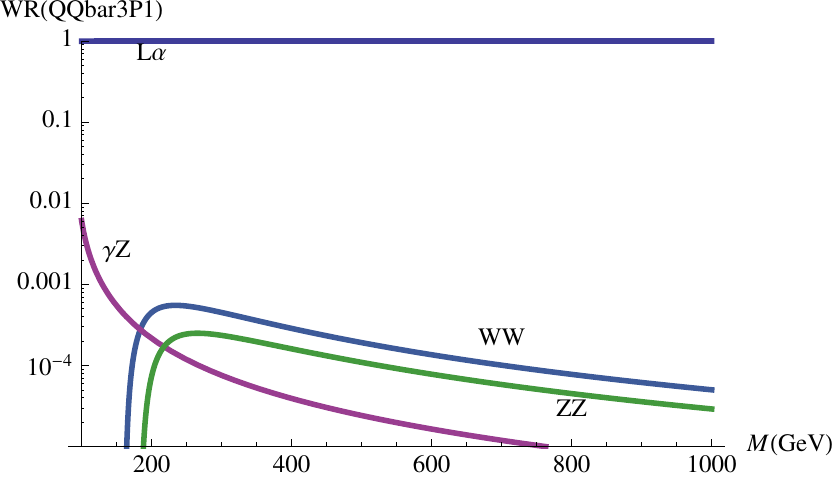}
		} 
	\subfloat[$^3P_2$]{
		\label{BR3P2nvector}
		\includegraphics[width=0.48\linewidth]{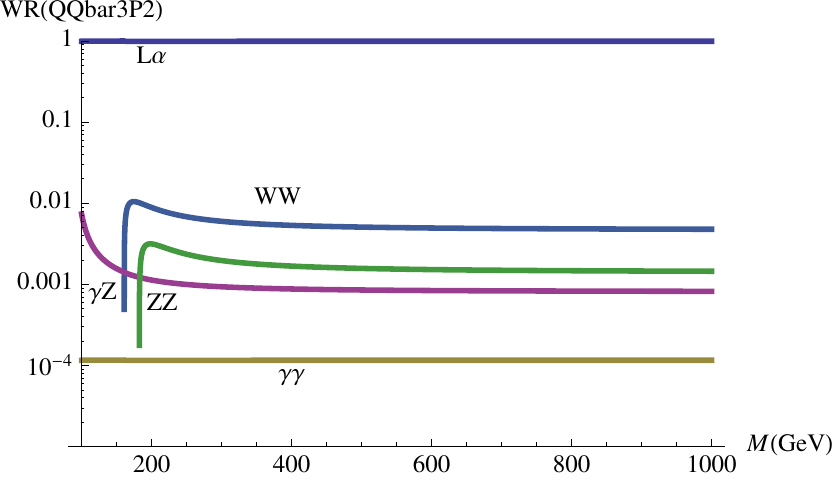}
		} \\

\caption{Decay width ratios of neutral vector-like quirkonia 
in different $J^{PC}$ states. 
Solid lines correspond to a Higgs mass $M_H = 125$ GeV, 
while dashed lines correspond to 
$M_H = 250$ GeV\@.  In many instances, there is no difference between
the width ratios for different Higgs masses, and thus the solid lines
overlap the invisible dashed lines.}
\label{BRSvector}
\end{figure*}

\begin{figure*}
	\centering
	\subfloat[$^3S_1$]{
		\label{BR3S1cvector}
		\includegraphics[width=0.48\linewidth]{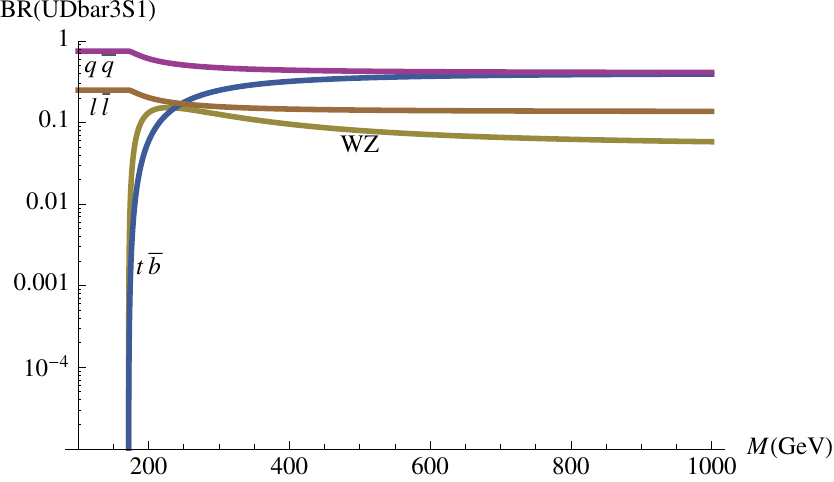}
		} 
	\subfloat[$^1P_1$]{
		\label{BR1P1cvector}
		\includegraphics[width=0.48\linewidth]{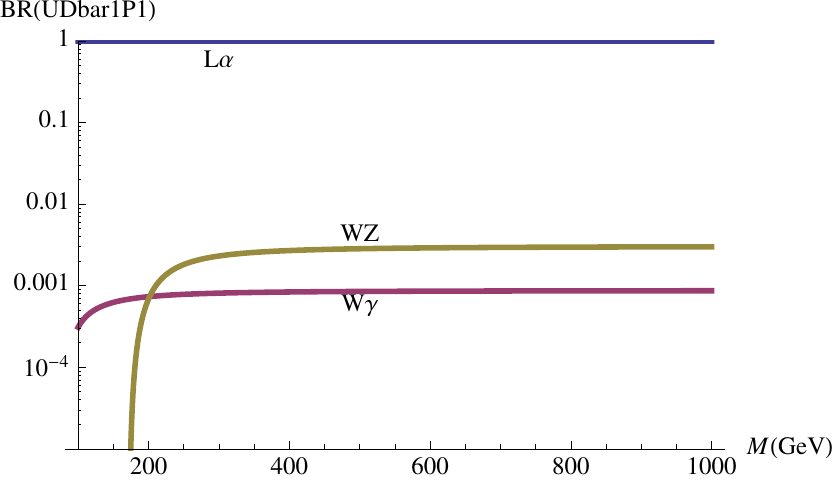}
		} \\

\caption{Same as Fig.~\ref{BRSvector} but for charged vector-like quirkonia. 
Only the two $J^{PC}$ states shown here have nontrivial branching ratios. 
The state $^3S_1$ cannot decay through two-body decays. 
The $^3P_{0,1,3}$ states can only decay radiatively into $S$ states.}
\label{BRScvector}
\end{figure*}

\section{Comparison to Vector-Like Quirkonia}
\label{sec:vectorlike}

Annihilation rates for the case of vector-like quirks in certain 
other representations has been calculated 
in \cite{Cheung:2008ke,Martin:2010kk}. 
There is not a general rule that relates the decay rates of 
vector-like quirks to chiral quirks. But in certain circumstances 
one can be obtained from the other, and vice versa. 
In this section, we will discuss differences and similarities of 
vector-like and chiral quirk decay rates, and give examples 
in cases where the decay rates are related.

We wish to compare our results for chiral quirks to a related theory 
with vector-like quirks.  The vector-like theory we consider consists 
of the doublet $Q$ given before in Table~\ref{table:quantumnumbers}, 
but now we replace
\begin{eqnarray}
\left[ \begin{array}{c} u^c(\bar{\mathbf{N}},\mathbf{1},-1/2) \\ 
                        d^c(\bar{\mathbf{N}},\mathbf{1},+1/2) 
       \end{array} \right] &\longrightarrow& 
Q'(\bar{\mathbf{N}}, \mathbf{2}, 0) \; .
\end{eqnarray}
Yukawa couplings, Eq.~(\ref{eq:yukawa}), are not present, and instead
we write the vector-like mass $M Q Q'$ where $M = M_U = M_D$.
There are several differences that lead to qualitatively 
different decay widths.

First, the coupling of electroweak gauge bosons to left- and 
right-handed quirks are the same -- the quirk-$W/Z$ coupling 
is a purely vector interaction, and processes that proceed through 
the axial vector coupling in the chiral case are absent for 
vector-like quirks. As an example, consider the decay rate 
$\Gamma (^3S_1 \rightarrow f\bar{f})$, for neutral and charged quirkonia. 
The only diagrams are the $s$-channel $\gamma/Z$ or $W$. 
In the neutral case, the only difference that separates 
vector-like and chiral is the different axial-vector and 
vector coupling of the $Z$. Therefore, the expressions for 
vector-like \cite{Cheung:2008ke} and chiral \cite{Barger:1987xg} 
are the same. For the charged case, the axial-vector and 
vector couplings are not explicitly written in \cite{Cheung:2008ke}, 
but their rate is 4 times larger than the chiral case 
in \cite{Barger:1987xg}. This is because the $s$-channel $W$ couples 
to both left and right handed quirks in the vector-like case, 
whereas in the chiral case they only couple to left handed quirks. 
Therefore, the decay rate into a fermion-antifermion pair 
for a charged $^3S_1$ is four times larger than its chiral counterpart.

Second, the quirks do not couple to the Higgs and the corresponding 
Goldstone bosons (through the longitudinally polarized electroweak 
gauge bosons).  Virtual Goldstone bosons can only appear in the
$s$-channel, and since the Goldstone bosons are pseudoscalars,
they only contribute to the $^1S_0$ decay rates.  Vector-like quirks,
by contrast, do not have couplings to the Higgs or the Goldstone bosons.
In addition, Goldstone bosons can appear in the final state 
(appearing as longitudinally polarized electroweak gauge bosons).
This leads to qualitatively different decay rates into gauge bosons 
for all of the bound states.

For completeness, we present the width and branching ratios of 
vector-like quirkonia in Fig.~\ref{BRSvector} for neutral quirkonia 
and Fig.~\ref{BRScvector} for charged. There are striking differences 
between the chiral and vector-like cases. The most prominent feature 
in the vector-like case is that all decay widths have the same 
asymptotic behavior at large quirkonium mass - there are no 
longitudinal enhancements of $W/Z$ anywhere. This is expected, 
as the longitudinal $W/Z$ asymptotes to the respective Goldstone bosons, 
which do not couple to the vector-like quirks in $u$- and $t$-channel 
quirk-exchange diagrams. Also, the trilinear gauge boson coupling 
appearing in $s$-channel  gauge boson exchange arises  
from the electroweak gauge structure of $SU(2)_L$ and has 
no relation to the electroweak breaking mechanism. 
Therefore, one would not expect any enhancements in the decay widths 
of vector-like quirkonia. Without longitudinal enhancements, 
the Lyman-alpha transition dominates over all $P$-state decays 
for all quirkonium masses. Whereas in the chiral case, 
decay channels that receives longitudinal enhancements can dominate 
the Lyman-alpha transition at large quirkonium masses. 

In the low quirkonium mass regime, the overall behavior of both 
vector-like and chiral quirkonia are similar:  $P$-states 
predominantly decay via the Lyman-alpha transition and 
$^3S_1$ into $q\bar{q}$. It is interesting to note that for 
$^1S_0$, $\gamma Z$ dominates the vector-like quirkonium decay, 
whereas $\gamma \gamma$ is dominant for chiral quirkonia. 
This is because the primordial electroweak gauge boson $W_3^{\mu}$ 
couples not just to the left-handed vector-like quirk, 
but to the right handed one also! A rough estimate indicates 
that this gives a factor of four increase in the $\gamma Z$ rate 
for the vector-like case. Indeed, the isospin contribution to the 
vector coupling of the $Z$ to the quirks for the vector-like case 
is twice as much as that for chiral quirks.

\section{Understanding Chiral Enhancements} 
\label{sec:chiralenhancement}

Figures \ref{BRc} and \ref{BRS} show that decay processes can be 
singly or doubly enhanced by either the Yukawa coupling or the 
longitudinal modes of gauge bosons at high quirkonium masses. 
A summary of enhancements received in the decay channels can be found 
in Tables~\ref{table:chargedenhancements} and \ref{table:neutralenhancements}. 
In this limit, the Goldstone equivalence theorem applies 
and the enhancements in various decay channels can be seen 
by the matching of the $J, P,$ and $C$ numbers between the 
decay quirkonium state and the final state consisting of 
Goldstone/Higgs bosons and/or transverse gauge bosons. 
First, we determine the $J^{PC}$ numbers to the final state particles; 
$0^{-+}$ and $0^{++}$ to the Goldstone boson and Higgs boson, respectively. 
For photons and transverse $W$ and $Z$, we assign $P$ and $C$ 
numbers according to the $P$ and $C$ of the bilinears 
$\bar{\psi} \gamma^{\mu} \psi$ and $\bar{\psi} \gamma^{\mu} \gamma^5 \psi$. 
The photon has $J^{PC}$ = $1^{--}$, and the $W/Z$ has $1^{--}$ 
for the vector coupling and $1^{++}$ for the axial vector coupling 
(that is, we absorb the violation of $C$ and $P$ from the axial vector 
coupling into the $J^{PC}$ of the $W$ and $Z$). 
For the quirkonia, we have 
$0^{-+}$ for $^1S_0$, 
$1^{--}$ for $^3S_1$, 
$1^{+-}$ for $^1P_1$, 
$0^{++}$ for $^3P_0$, 
$1^{++}$ for $^3P_1$ and finally, 
$2^{++}$ for $^3P_2$.

\begin{table}[t]
\begin{tabular}{|c|c|c|}
\hline
$J^{PC}$ & Singly enhanced        & Doubly enhanced \\ \hline
$^1S_0$  &                        & $WH$ \\
$^3S_1$  & $W\gamma$, $WH$, $WZ$  & \\
$^1P_1$  & $W\gamma$, $WH$, $WZ$  & \\
$^3P_0$  &                        & \\
$^3P_1$  & $W\gamma$, $WZ$        & $WH$ \\
$^3P_2$  & $W\gamma$, $WH$, $WZ$  & \\
\hline
\end{tabular}
\caption{Decay channels that receive enhancements for charged quirkonia.}
\label{table:chargedenhancements}
\end{table}

\begin{table}[t]
\begin{tabular}{|c|c|c|}
\hline
$J^{PC}$ & Singly enhanced                         & Doubly enhanced \\ \hline
$^1S_0$  &                                         &  $ZH$     \\
$^3S_1$  & $Z\gamma$,$ZZ$, $ZH$, $\gamma H$        & $WW$  \\
$^1P_1$  & $Z\gamma$, $WW$, $ZZ$, $ZH$, $\gamma H$ & \\
$^3P_0$  &                                         & $WW$, $ZZ$, $HH$ \\
$^3P_1$  & $Z\gamma$, $WW$, $ZZ$                   & $ZH$ \\
$^3P_2$  & $Z\gamma$, $ZH$                         & $WW$, $ZZ$, $HH$ \\
\hline
\end{tabular}
\caption{Decay channels that receive enhancements for neutral quirkonia.}
\label{table:neutralenhancements}
\end{table}

Next, we write down all the available $J^{PC}$ with different 
orbital angular momentum $L$ between the two final state particles, 
and match with the $J^{PC}$ of the quirky meson to determine 
which meson decay channels are enhanced. As an example, 
consider $\bar{Q}Q \to Z\gamma$, where there are only the $t$- and 
$u$-channel diagrams. Single enhancement from the longitudinal $Z$ 
is present in some meson states. In the limit where the quirkonium mass 
$M \gg M_Z$, the longitudinal $Z$ is equivalent to the corresponding 
Goldstone boson $\phi^0$. The $J^{PC}$ of the $\phi^0\gamma$ system 
is determined by combining that of the Goldstone boson $(0^{-+})$ 
and that of the photon $1^{--}$, which gives $1^{+-}$. 
Keeping the total angular momentum $J \leq 2$, we can add orbital 
angular momentum $L$ into the system, forming 
$\{0^{--}, 1^{--}, 2^{--}\}$ for $L=1$, and 
$\{1^{+-}, 2^{+-}\}$ for $L=2$. 
This is captured in Table~\ref{table:JPCphigamma}. 
One sees that, only the states $1^{+-}$ and $1^{--}$ matches with 
the existing quirky meson states $^1P_1$ and $^3S_1$, respectively. 
One can see from Appendix~\ref{Zgamma}, that only the 
$^1P_1$ and $^3S_1$ states are enhanced. One can also see that both 
of the decay proceed via the axial vector coupling of the $Z$. 
This is because the $\phi^0 \bar{f}f$ coupling is a 
pseudoscalar coupling $\bar{\psi} \gamma^5 \psi$, and this must 
correspond to the axial vector coupling of the $Z$, 
$\bar{\psi} \gamma^{\mu}\gamma^5 \psi$.

\begin{table}[t]
\begin{tabular}{c|ccc}
       & $J = 0$  & $J=1$    & $J=2$    \\ \hline
$L=0$  &          & $1^{+-}$ &          \\
$L=1$  & $0^{--}$ & $1^{--}$ & $2^{--}$ \\
$L=2$  &          & $1^{+-}$ & $2^{+-}$ \\
\end{tabular}
\caption{$J^{PC}$ of the $\phi^0 \gamma$ system.}
\label{table:JPCphigamma}
\end{table}

We will go further and illustrate that all other meson states decay 
to $Z\gamma$ via the vector coupling and receive no enhancements 
with the same procedure. Consider the final state with a transverse 
$Z$ and $\gamma$, both with $J^{PC}$ = $1^{--}$ for the vector coupling 
of the $Z$. The $J^{PC}$ of the final states are collected in 
Table~\ref{table:JPCZgamma}, where one can see that the 
$^1S_0, ^3P_0, ^3P_1$, and the $^3P_2$ states are not enhanced. 
The $J^{PC}$ of the final states with an axial vector coupling can be 
obtained by flipping the $C$ and $P$ numbers everywhere in 
Table~\ref{table:JPCZgamma}. Then one sees that there are also 
contributions with no longitudinal $Z$ enhancements via the axial vector 
coupling to the decay widths of $^3S_1$ and $^1P_1$. 
Indeed, there are terms in the expressions for the corresponding 
decay widths that are not enhanced.

\begin{table}[t]
\begin{tabular}{c|ccc}
      &$ J = 0$        & $J=1$    & $J=2$ \\ \hline
$L=0$   & $0^{++}$  & $1^{++}$ & $2^{++}$        \\
$L=1$ & $0^{-+}$& $1^{-+}$ &   $2^{-+}$ \\
$L=2$ &  &$1^{++}$&   $2^{++}$ \\
\end{tabular}
\caption{$J^{PC}$ of the $Z_T \gamma$ system with vector coupling of the $Z$.}
\label{table:JPCZgamma}
\end{table}

The procedure above can explain a large number of enhancements 
for different quirkonium decay processes. However, there are 
instances where the procedure predicts leading enhancements 
in some processes when there should not have been any. 
The fictitious leading enhancements predicted by this procedures are: 
$^1S_0 \to W_T^-\phi^+$, 
$^1S_0 \to \phi^+ \gamma$, 
$^1P_1 \to \phi^+ H$, 
$^3P_0 \to W_T^+ H$ and 
$\{^1S_0$, $^3P_0\} \to Z_T \phi^+$.

The $^1S_0$ state are not singly enhanced in two-body systems 
consisting of one transverse gauge boson. This can be seen by 
tracking the spin along the quirky fermion line. 
First, the $^1S_0$ projector is proportional to 
$\sum_{s_1 s_2} C_{^1S_0}^{s_1 s_2} u_{s_1} \bar{v}_{s_2}$, 
where $C_{^1S_0}^{s_1 s_2}$ is the Clebsh-Gordon coefficients corresponding 
to the $^1S_0$ state, which vanishes for $s_1 =  s_2$. 
In the systems being considered, there is only one vertex that 
flips the spin - the fermion-fermion-gauge boson vertex. 
The rest are either scalar (Higgs boson) or pseudoscalar (Goldstone boson) 
couplings that do not involve a spin flip. Also, one can see that 
the Dirac spinor $v$ has an opposite spin compared to the 
Dirac spinor $u$ in the relation $v_s = -2s \gamma^5 u_{-s}$. 
Therefore, the $^1S_0$ state cannot decay into any states consisting 
of only one transverse gauge boson (with the other outgoing particle 
being the Higgs or Goldstone boson) and the $^1S_0$ decay width 
cannot be singly enhanced.

\section{Discussion}
\label{sec:discussion}

We have calculated the chiral quirkonium decay rates for quirks 
that acquire mass through interactions with the Higgs boson.
While in this paper we have not studied the production rates of 
quirks at the LHC, this is straightforward using standard collider
physics tools, and has been done recently in the literature.  
For vector-like quirks with electroweak quantum numbers, 
the production cross section was calculated in Fig.~6 of 
Ref.~\cite{Martin:2010kk}.  There it was shown that pair production 
of uncolored quirks produced through $\gamma$/$Z$ exchange has 
a cross section ranging from roughly $\sim 10^5$ to $\sim 10$~fb 
at $\sqrt{s} = 7$ TeV LHC for the quirkonium mass range of 
$100$ to $1000$~GeV\@.  We expect that the chiral quirk production 
cross sections are very similar in size. 
Given the spectacular signals that result from quirkonia decay, 
the discovery of quirkonium resonances can occur quickly.
This is likely to occur well before the detailed properties 
(spin, $P$, $C$, etc.) of the resonances can be determined.


The particular decay channels not only can tell us about
the constituent quirks' quantum numbers, but perhaps even more
interestingly, how the quirks acquire mass.  
Quirkonia with chiral quirks have longitudinal and Yukawa enhancements 
that are absent or highly suppressed in quirkonia with vector-like
quirks.  In this paper we have demonstrated the striking differences
between the dominant chiral quirkonium decay channels as compared 
with vector-like quirkonium decay channels.  This should enable 
the LHC to easily distinguish whether quirks are chiral or vector-like 
from the observation and branching ratios of the dominant decay channels.

For electrically-charged chiral quirkonia composed of quirks with 
the quantum numbers given in this paper, the state $^1S_0$ decays 
into $WH$ or $t \bar{b} / \bar{t} b$ overwhelmingly for quirkonium 
masses larger than about $250$~GeV\@.  Contrast this with 
vector-like quirkonia, where the $^1S_0$ with the quantum numbers
given earlier in the paper, does not even have two-body decays. 
Chiral quirkonia in the $^3S_1$ state have $WZ$ is the dominant 
decay channel.  For vector-like quirkonia, the $f \bar{f}$, summed
over all flavors of SM fermions, dominates. 
We also demonstrated that the Lyman-alpha transition is dominant 
in all of the charged quirkonia $P$-states, 
except for $^3P_1$ for chiral quirkonia 
when the ``doubly-enhanced'' $WH$ decay becomes significant for 
quirkonium masses $\gtrsim 600$ GeV\@.  The $W\gamma$ decay deserves 
more discussion.  It was shown in \cite{Burdman:2008ek} that the 
$W\gamma$ channel is dominant when their squirk and anti-squirk pair 
has low relative velocity, in other words, an $S$-state. 
For our case, the $W\gamma$ partial width vanishes due to our 
choice of quantum numbers:  $Q_U = -Q_D = 1/2$, where we found 
the $W\gamma$ partial width is proportional to $(Q_U + Q_D)^2$, 
see Appendix~\ref{subsec:Wgamma}.

For electrically-neutral chiral quirkonia, the dominant decay channels 
of $^1S_0$ are $ZH$ and $t\bar{t}$.  Again, contrast this with 
vector-like quirkonia where $WW$ or $ZZ$ dominates.  
For $^3S_1$, the $WW$ channel dominates for chiral quirkonia, 
versus $f\bar{f}$ for vector-like quirkonia.  For all $P$-states, 
Lyman-alpha emission dominates for vector-like quirkonia, whereas for 
chiral quirkonia there are several decay channels that can become 
significant when the quirkonium mass is large.  In particular, 
$HH$ can dominate for $^3P_0$, $ZH$ for $^3P_1$, and $WW$ for $^3P_2$.

Indeed, perhaps one of the most interesting decay channels 
that we found is the $^3P_0$ decay into two Higgs bosons, 
which becomes the dominant decay channel for 
quirkonium masses $\gtrsim 500$ GeV\@.  This could be striking
signal at LHC, given that the di-Higgs system would reconstruct
to an invariant mass peak of the $^3P_0$ state.

Finally, it is tempting to consider applications of our results to 
various existing hints at colliders.  For instance, the prominence of the
electrically-charged chiral quirkonium decay channel, $^1S_0 \ra WH$
is suggestive:  chiral quirkonia with mass $M \sim 300$ GeV with
some minor modification of the $H$ decay into jets could easily 
lead to the CDF excess in the $Wjj$ 
signal \cite{Aaltonen:2011mk,CDFWjjwebsite}.
The cross section can be easily adjusted to match the excess, 
simply by enlarging the number of infracolors or flavors of quirks.
Given the incredible performance of the LHC over the past several
months, we leave this pursuit to future work.

\section*{Acknowledgments}

We thank Z.~Chacko, R.~Harnik, and A.~Martin for many useful discussions.
GDK was supported by a Ben Lee Fellowship from Fermilab.
RF and GDK were supported in part by the US Department of Energy 
under contract number DE-FG02-96ER40969 and by NSF 
under contract PHY-0918108.
Fermilab is operated by Fermi Research Alliance, LLC, 
under Contract DE-AC02-07CH11359 with the US Department of Energy.


\appendix

\begin{widetext}

\section{Matrix elements of bound state decays}
\label{sec:me}

This section reviews the procedures to evaluate the decay amplitudes 
of different angular momentum bound states following  the method 
in \cite{Guberina:1980dc}. We work in the non-relativistic limit, 
where the relative momentum of the constituents, $|\mathbf{q}| \ll M$, 
where M is the mass of the meson. We also ignore the contribution 
to the meson mass from the binding potential, i.e., we take $M = 2m_Q$, 
with $m_Q$ the mass of the individual quirks. 

Calculations of the matrix element involving an incoming bound state 
and an outgoing free state, $\langle X|iT|B\rangle$, are needed 
to evaluate different bound state decay rates.  This is most conveniently 
done by writing the bound state as a superposition of free fermion states 
with spins $(s_1,s_2)$ and momenta $(p_1,p_2)$:

\begin{eqnarray}
|B\rangle=|^{2s+1}l_j\rangle 
  &=& \sum_{M S_z} \langle l m s s_z | j j_z \rangle |l m s s_z \rangle 
      \nonumber \\
  &=& \sqrt{\frac{2}{M}} \int \!\!\! \frac{d^3 \mathbf{q}}{(2\pi)^3} 
      \sum_{m s_z} \psi^{lm}(\mathbf{q})
      \langle l m s s_z | j j_z \rangle \times 
      \nonumber \\ 
  & & \bigg[ \sum_{s_1 s_2}
             \langle s_1, \frac{1}{2}, s_2,  \frac{1}{2} | s s_z \rangle 
             \bigg] |s_1p_1 s_2 p_2 \rangle,
\end{eqnarray}
where $\psi$ is the Schr\"odinger wavefunction of the bound state. 
In its rest frame, $p_1 = Q/2 + q$, and $p_2 = Q/2 - q$, where $Q$ 
is the 4-momentum of the meson, and $q$ is the relative 4-momentum 
of quirks. Then, the quantity 
$<\!\!X | iT |s_1p_1 s_2 p_2\!\!> = i \bar{v}_{s_2}(p_2) 
\mathscr{M} u_{s_1}(p_1)$ is the usual fermion-antifermion 
annihilation matrix element into the outgoing state $f$. 
Expanding the above to the lowest non-vanishing order in $\mathbf{q}$, 
we found the following decay amplitudes for S and P states,
\begin{eqnarray}
A(^1S_0) &=& \sqrt{\frac{\N}{16 \pi M}} R_S(0)\,
              Tr[\mathscr{M} \gamma^5 (-Q\!\!\!\!/\, + M)], \label{decayamp} \\
A(^3S_1) &=& \sqrt{\frac{\N}{16 \pi M}} R_S(0)\,
             Tr[\mathscr{M} \epsilon\!\!\!/\,  (-Q\!\!\!\!/\, + M)], \\
A(^1P_1) &=& -i\sqrt{\frac{3 \N}{4 \pi M}} R_P'(0)\,
             Tr[\frac{1}{2}\epsilon_{\mu}\mathscr{M}^{\mu} \gamma^5(-Q\!\!\!\!/\, + M) + \mathscr{M} \epsilon\!\!\!/\, \frac{Q\!\!\!\!/\,}{M} \gamma^5],\\
A(^3P_0) &=& i\sqrt{\frac{\N}{4 \pi M}} R_P'(0)\,
             Tr[\frac{1}{2}\mathscr{M}^{\alpha}\bigg(\frac{Q_{\alpha}Q\!\!\!\!/\,}{M^2}-\gamma_{\alpha}\bigg)(-Q\!\!\!\!/\, + M) - 3\mathscr{M}], \\
A(^3P_1) &=& i \sqrt{\frac{3 \N}{8 \pi M}}R_P'(0) 
             Tr[2 \mathscr{M} \epsilon\!\!\!/\, \gamma^5 - \frac{i}{2M} \epsilon^{\rho \alpha \beta \delta}Q_{\rho} \mathscr{M}_{\alpha}\gamma_{\beta} \epsilon_{\delta} (-Q\!\!\!\!/\, + M)], \quad \textrm{and} \\
A(^3P_2) &=& -i \sqrt{\frac{3 \N}{4 \pi M}}R_P'(0) 
             Tr[\frac{1}{2}\mathscr{M}_{\alpha} \epsilon^{\alpha \beta} \gamma_{\beta}(-Q\!\!\!\!/\, + M) ],
\label{decayamp2}
\end{eqnarray}
where $R(0)$ and $R'(0)$ are the meson radial wavefunction and its derivative 
at the origin, respectively.  

The mass of each meson is distinct, and in principle the $M$ in
each of the above expressions should be replaced with the mass
for that particular bound state.  Since we assume $\bar{\alpha}_{ic}$
is perturbative, the differences between bound state energies
is parametrically small, of order $\bar{\alpha}_{ic}^2 M$.  
In most instances, one can use our expressions below, substituting
the proper quirkonia mass for $M$, and computing the rates.
For the branching ratio plots we present below, however, this
difference is small.  

The quantity 
$\mathscr{M}_{\alpha} = \partial\mathscr{M} / \partial q^{\alpha}$ 
is the derivative of the matrix element with respect to the 
relative momentum $q$. The meson polarizations in the rest frame, 
$\epsilon^{\mu}$ for spin-1, and $\epsilon^{\mu \nu}$ for spin-2, 
are chosen to be
\begin{eqnarray}
\epsilon^{\mu}_{\mp} &=& (0,\mp \frac{1}{\sqrt{2}}, - \frac{i}{\sqrt{2}} , 0),
                         \nonumber \\
\epsilon^{\mu}_{L}   &=& (0,0,0,1), \nonumber \\
\epsilon^{\alpha\beta}_{J_z} &=& \sum_{M,S_z} \langle 1M, 1S_z| 2J_z \rangle,
\end{eqnarray}
with $J_z \in \{-2,-1,0,1,2\}$. The factor of $\sqrt{N}$ arises from 
normalizing the meson as an infracolor singlet, exactly analogous
to what is done with the QCD color factor for quarkonia \cite{Barger:1987xg}.
 
The $P$ and $C$ parities of the above angular momentum states 
are manifest in each of the decay amplitudes above. 
For example, with $Q^{\mu} = (M,0,0,0)$ in the rest frame, 
the $P$ and $C$ parities of the bilinear constructed from the 
projector appearing in 
$A(^1S_0)$, $ \bar{\psi} \gamma^5 (-Q\!\!\!\!/\, + M) \gamma^0 \psi$, 
are $-$ and $+$, respectively. Thus $J^{PC} = 0^{-+}$ for $^1S_0$, 
as expected. One can check that the other amplitudes give the 
expected $J^{PC}$ using the same procedure. 
From Eqs.~(\ref{decayamp}) to (\ref{decayamp2}), we rederived all 
of the two-body decay rates listed in \cite{Barger:1987xg}.\footnote{We 
found a relative sign difference between the two terms in the 
amplitude $A(^3P_1)$. We attribute this to our definition of 
$\epsilon^{0123} = 1$.}

\section{Decay Rates of Charged Quirkonia}
\label{sec:chargedquirkoniumdecays}

Apart from a color factor of $\N$, decay rates of neutral quirkonia 
that do not involve any gluons are the same as listed in 
\cite{Barger:1987xg}. The decay rates of charged quirkonia will be 
discussed in this section. 

Charged quirkonia are expected to have larger partial widths 
than their neutral counterparts. This is because charged particles 
do not have a well-defined charge conjugate parity, hence loosening 
the constraints posed by $CP$ conservation. Here, we list the 
partial widths of charged quirky mesons with positive unit electric charge, 
i.e. $Q_U-Q_D = 1$, where Q is the electric charge of either the 
up-type or down-type quirks. The mass ratio squared $R_i$ and the 
relative velocity $\beta_{i,f}$ appearing in the formulas 
below are defined as
\begin{eqnarray}
R_i &=& \frac{m_i^2}{M^2}, \quad \quad \textrm{and} \\
\beta_{i,j} &=& \sqrt{1+ (R_i - R_j)^2 + 2(R_i +R_j)},
\end{eqnarray}
respectively, and $c_W = \cos \theta_W$ is the cosine of the Weinberg angle.
 
\subsection{$W^+ \gamma$}
\label{subsec:Wgamma}

The charged quirkonium decay widths into $W^+ \gamma$ are qualitatively 
different to the widths of neutral quirkonia into $Z\gamma$. 
There are two reasons for the differences; the decay into $W^+ \gamma$ 
can go through an $s$-channel with a $W$ exchange. The corresponding diagram 
is absent for $Z\gamma$; the photon does not couple to a 
electrically neutral $Z$. Another reason is that the photon couples 
to quirks of different electric charges in the $t$- and $u$-channel diagrams, 
due to the emission of a charged $W$. It is illuminating to write 
the chiral projection operators as $P_{L,R} = (v_W \mp a_W \gamma^5)/2$, 
with $a_W = v_W =1$, so that the vector and axial-vector contributions 
from the $W$ current are manifest. The partial widths into $W^+ \gamma$ are
\begin{eqnarray}
\Gamma(^1S_0^+ \to W^+ \gamma) & = &  \frac{\N \alpha \alpha_W v_W^2}{4M^2}(Q_U+Q_D)^2(1-R_W)|R_S(0)|^2, \\
\Gamma(^3S_1^+ \to W^+ \gamma) &  = & \frac{\N \alpha \alpha_W a_W^2}{12m_W^2}(Q_U+Q_D)^2 (1-R_W^2) |R_S(0)|^2, \\
\Gamma(^1P_1^+ \to W^+ \gamma) &  = & \frac{\N\alpha \alpha_W}{M^2 m_W^2} [a_W^2 (Q_U+Q_D)^2(1-R_W^2)+v_W^2(Q_U-Q_D)^2R_W(1-R_W)]|R_P'(0)|^2, \\
\Gamma(^3P_0^+ \to W^+ \gamma) &  = & \frac{\N \alpha \alpha_W (1-R_W)}{M^4}\bigg[ a_W^2(Q_U-Q_D)^2 + v_W^2(Q_U+Q_D)^2\bigg(1 + \frac{2}{1-R_W}\bigg)^2\bigg] |R_P'(0)|^2, \\
\Gamma(^3P_1^+ \to W^+ \gamma) & = &  \frac{\N \alpha \alpha_W}{2M^2 m_W^2} \bigg[a_W^2(1-R_W) + 4v_W^2(Q_U+Q_D)^2R_W^2\frac{1+R_W}{1-R_W}\bigg] |R_P'(0)|^2, \\
\Gamma(^3P_2^+ \to W^+ \gamma) & = &\frac{\N \alpha \alpha_W(1-R_W)}{10M^2 m_W^2}\bigg[ a_W^2(Q_U-Q_D)^2(3+4R_W) + \frac{4v_W^2 (Q_U+Q_D)^2 R_W (6 + 3R_W + R_W^2)}{(1-R_W)^2}\bigg] |R_P'(0)|^2. \nonumber \\
\end{eqnarray}
Interestingly, all but one term in $\Gamma(^3P_1^+ \to W^+ \gamma)$ 
are proportional to either the hypercharge $Y = (Q_U+Q_D)/2$ 
or the isospin $T_{3U} = (Q_U - Q_D)/2$ of the quirks.
\subsection{$W^+ H$}

In the limit degenerate quirk masses, their coupling constants 
to the Higgs boson are the same. As a consequence, the decay matrix 
elements has the same form as that for the decay into $ZH$, 
and can be obtained by the replacements $g_Z \to g/\sqrt{2}$ 
and vector and axial vector couplings by $1/2$; $a, v \to 1/2$. 
This gives a conversion factor of $1/(2\sqrt{2})$ converting the 
$ZH$ matrix elements to $WH$:
\begin{equation}
\mathscr{M}_{W^+H} = \frac{1}{2\sqrt{2}}\mathscr{M}_{ZH}.
\end{equation}
Therefore, the partial widths into $W^+ H$, have exactly the 
same form as those for $ZH$, aside from a factor of $1/8$. 
The analysis for $ZH$ in \cite{Barger:1987xg} applies to $W^+H$ 
as well. The partial widths are
\begin{eqnarray}
\Gamma(^1S_0^+ \to W^+ H) & = &  \frac{\N \alpha_W^2 \beta_{WH}^3}{32 M^2}\frac{1}{ R_W^2} |R_S(0)|^2, \\
\Gamma(^3S_1^+ \to W^+ H) & = & \frac{\N \alpha_W^2 \beta_{WH}}{384}\frac{M^2}{m_W^4} \bigg(\frac{8 R_W [(1 - R_W)^2 + R_H (1 - 3 R_W)]^2}{(1-R_W)^2(1-R_H-R_W)^2} \nonumber \\
&&+ \frac{[R_H^2 (1 - 3 R_W) -
  2 R_H (1 - R_W (2 + R_W)) + (1 - R_W) (1 - R_W^2 - \beta_{WH}^2)]^2}{(1-R_W)^2(1-R_H-R_W)^2}\bigg)|R_S(0)|^2 ,\nonumber \\
  & & \\
\Gamma(^1P_1^+ \to W^+ H) &= & \frac{\N \alpha_W^2 \beta_{WH}^3}{4M^2 m_W^2(1-R_H-R_W)^2}|R_P'(0)|^2,\\
\Gamma(^3P_0^+ \to W^+ H) &= & 0, \\
\Gamma(^3P_1^+ \to W^+ H) & = &\frac{\N \alpha_W^2 \beta_{WH}}{8 m_W^4} \bigg( \frac{2[1-R_H+R_W]^2[1+R_W(2-R_H+R_W)]^2}{(1-R_H-R_W)^2(1-R_W)^2} \nonumber \\
& & + R_W \bigg[ \frac{4R_W}{1-R_W} + \frac{\beta_{WH}^2-4(1-R_H-R_W)}{(1-R_H-R_W)^2}\bigg]^2\bigg) |R_P'(0)|^2,\\
\Gamma(^3P_2^+ \to W^+ H)& = & \frac{3 \N \alpha_W^2 \beta_{WH}^5}{40 M^2m_W^2(1-R_H-R_W)^4} |R_P'(0)|^2.
\end{eqnarray}

\subsection{$W Z$}

Notice that double longitudinal modes are allowed from the decay 
of a charged quirkonium in the $^1S_0$ state. This is impossible 
for the neutral quirkonium case, where it decays into $ZZ$ or $WW$. 
To see this, the $^1S_0$ state has $J^{PC}=0^{-+}$, but at 
zero angular momentum, the double longitudinal state has $J^{PC}=0^{++}$. 
The decay into double longitudinal modes for neutral quirkonia in $^1S_0$ 
is forbidden by $CP$ conservation. For charged states, the charge parity 
is irrelevant, and the decay into double longitudinal mode is 
allowed by $CP$ conservation. Naively, one would expect the $^1S_0$ 
decay rate is longitudinal from appearance of the $1/(R_Z R_W)$ term. 
However, due to the Goldstone s-exchange, at large quirkonium mass 
$M$ the decay rate vanishes.

\begin{eqnarray}
\Gamma(^1S_0^+ \to W^+ Z) & = & \frac{\N \alpha_W \alpha_Z \beta_{WZ} }{32 M^2} \Bigg( 1 - \frac{c^2_{W} R_Z}{R_W} \frac{1}{1+R_W-R_Z}\Bigg)^2 \Bigg(\frac{8}{(1-R_W-R_Z)^2}+ \frac{1}{R_W R_Z}\Bigg)  |R_S(0)|^2, \nonumber \\
& &  \\
\Gamma(^3S_1^+ \to W^+ Z) &= &\frac{\N \alpha_W \alpha_Z \beta_{WZ}^3}{64 M^2}\frac{1}{ (1-R_W)^2(1-R_W-R_Z)^2} \bigg\{8c_W^4 R_Z^2 \nonumber \\
&& + 2(1-R_W-2c_W^2R_Z)^2\bigg(\frac{1}{R_W}+\frac{1}{R_Z}\bigg) \nonumber \\
&& + \frac{1}{R_W R_Z}(1-R_W-c_W^2 R_Z(1+R_W +R_Z))^2\bigg\} |R_S(0)|^2,\\
\Gamma(^1P_1^+ \to W^+ Z) & = &\frac{\N \alpha_W \alpha_Z \beta_{WZ}}{4M^4(1-R_W-R_Z)^2}\bigg\{\frac{(1+R_W-R_Z)^2}{R_W}+\frac{(1-R_W+R_Z)^2}{R_Z} \nonumber \\
&& +4\bigg(1-\frac{c_W^2 \beta_{WZ}^2}{1-R_W-R_Z}\bigg)^2\bigg\} |R_P'(0)|^2, \\
\Gamma(^3P_0^+ \to W^+ Z) & = & \frac{\N \alpha_W \alpha_Z \beta_{WZ}^3}{M^4(1-R_W-R_Z)^4}[1-c_W^2(1-R_W+R_Z)]^2 |R_P'(0)|^2,\\
\Gamma(^3P_1^+ \to W^+ Z) &= & \frac{\N \alpha_W \alpha_Z \beta_{WZ}^3}{16M^4(1-R_W-R_Z)^2} \bigg\{\frac{32 c_W^4 R_Z^2}{(1-R_W)^2} \nonumber \\
&& + \frac{2}{R_Z} \bigg[ 1+\frac{2R_Z}{1-R_W-R_Z} - \frac{8 c_W^2 R_Z \big(1- \frac{R_Z}{2(1-R_W)}\big)}{1-R_W-R_Z}\bigg]^2 \nonumber \\
&& +\frac{2}{R_W} \bigg[  2c_W^4 R_Z \bigg(1+\frac{2R_W+R_Z}{1-R_W}\bigg)^2  + \bigg( 1+\frac{2R_W}{1-R_W-R_Z} -2c_W^2 (1-\frac{2R_Z}{1-R_W})\bigg)^2\bigg]  \nonumber \\
&& \bigg\} |R_P'(0)|^2\\
\Gamma(^3P_2^+ \to W^+ Z) &= & \frac{\N \alpha_W \alpha_Z \beta_{WZ}^3}{40M^4(1-R_W-R_Z)^4} \bigg\{ 16[1-c_W^2(1-R_W+R_Z)]^2 \nonumber \\
&& +\frac{3}{R_Z}[1-R_W+R_Z -4c_W^2 R_Z]^2 \nonumber \\
&& +\frac{3}{R_W}[1+R_W-R_Z -2c_W^2(1-R_W-R_Z)]^2\bigg\}|R_P'(0)|^2. \nonumber \\
\end{eqnarray}
\subsection{$f_u\bar{f}_d$}
\label{ffbar}

Decays into two fermions only proceed via the $s$-channel exchange of $W^+$. 
The non-zero widths with outgoing fermion masses $m_{1,2}$ are
\begin{eqnarray}
\Gamma(^1S_0^+ \to u \bar{d}) & = & \frac{\N \alpha_W^2 \beta_{ud}}{16 M^2} \frac{(R_1 - R_1^2 + R_2 + 2 R_1 R_2 - R_2^2)}{R_W^2} |R_S(0)|^2, \\
\Gamma(^3S_1^+ \to u \bar{d}) & = & \frac{\N \alpha_W^2 \beta_{ud}}{48 M^2} \frac{2 - R_1 - R_1^2 - R_2 + 2 R_1 R_2 - R_2^2}{(1-R_W)^2} |R_S(0)|^2,\\
\Gamma(^3P_1^+ \to u \bar{d}) & = & \frac{\N \alpha_W^2 \beta_{ud}}{2M^4}\frac{2 - R_1 - R_1^2 - R_2 + 2 R_1 R_2 - R_2^2}{(1-R_W)^2} |R_P'(0)|^2.
\end{eqnarray}
As expected, the $^1S_0$ partial width is proportional to $m_f^2/M^2$, 
corresponding to a chirality flip on the outgoing fermion line.


\section{Decay Rates of Neutral Quirkonia}
\label{sec:neutralquirkoniumdecays}

This section summarizes the decay rates of neutral quirkonia. 
The decay rates differ with \cite{Barger:1987xg} by just a factor of $2/3$ 
due to a different color group. We also attempt to rewrite the 
decay rates so that the origins of the terms in the expression 
are manifest. In the expressions below, a $t$-channel quirk exchange 
with outgoing particles $i$ and $j$ corresponds to the factor 
$(1-R_i -R_j)^{-1}$, with $R_i = m_i^2/M^2$, and $M$ is the quirkonium mass. 
An $s$-channel diagram exchanging particle $\phi$ corresponds to 
$(1-R_{\phi})^{-1}$.

\subsection{$f \bar{f}$}

The decays into a fermion-antifermion pair, only the $s$-channel 
$\gamma, Z$ and Higgs diagram contribute. The decay of the $^1S_0$ state 
requires a chirality flip on the outgoing fermion line, 
resulting in the dependence on the fermion mass squared, $M_f^2$, 
in its decay rate - similar to pseudoscalar decay.

\begin{eqnarray}
\Gamma(^1S_0 \to f \bar{f}) & = &  8\N \alpha_Z^2 a_f^2 a_Q^2 \beta_f\frac{m_f^2}{m_Z^4}|R_S(0)|^2, \\
\Gamma(^3S_1 \to f \bar{f}) &  = & \frac{4\N \alpha_{EM}^2 \beta_f}{3M^2}\bigg\{ (1+2R_f)\bigg(e_Q e_f  +\frac{v_f v_Q}{c_W^2 s_W^2 (1-R_Z)}\bigg)^2 + \frac{a_f^2 v_Q^2\beta_f^2}{c_W^4 s_W^4 (1-R_Z)^2}\bigg\} |R_S(0)|^2, \\
\Gamma(^1P_1 \to f \bar{f}) &  = & 0, \\
\Gamma(^3P_0 \to f \bar{f}) &  = & \frac{9 \N \alpha_Z^2 \beta_f^3}{8M^2(1-R_H)^2} \frac{m_f^2}{m_Z^4}|R_P'(0)|^2, \\
\Gamma(^3P_1 \to f \bar{f}) & = & \frac{32\N \alpha_Z^2 a_Q^2 \beta_f}{M^4(1-R_Z)^2}[a_f^2 \beta_f^2 + (1+2R_f)v_f^2] |R_P'(0)|^2, \\
\Gamma(^3P_2 \to f \bar{f}) & = & 0, \label{eq:ffbar}
\end{eqnarray}

where $M$ is the quirkonium mass, 
$\alpha_Z = \alpha_{EM} / (c_W^2 s_W^2)$, $c_W$ and $s_W$ are the 
cosine and sine of the Weinberg angle, respectively, 
$a_i = T_{3i}/2$ and $v_i = a_i - e_i s_W^2$ are the axial-vector 
and vector couplings of the $Z$ to fermion $i$, with $i = \{ Q,f\}$ 
for the quirk and the outgoing fermion, respectively, $R_j = m_j^2/M^2$, 
and $\beta_f = \sqrt{1-4R_f}$ is the relative velocity between the 
two outgoing fermions.

\subsection{$Z\gamma$}

Only the $t$-channel diagram contributes decays into $Z\gamma$,

\begin{eqnarray} 
\Gamma(^1S_0 \to Z\gamma) & = & \frac{8 \N \alpha_{EM}\alpha_Z e_Q^2 v_Q^2}{M^2}(1-R_Z)|R_S(0)|^2, \\
\Gamma(^3S_1 \to Z\gamma) &  = & \frac{8\N \alpha_{EM}\alpha_Z  e_Q^2 a_Q^2}{3 m_Z^2} (1-R_Z^2)|R_S(0)|^2, \\
\Gamma(^1P_1 \to Z\gamma) &  = & \frac{32\N \alpha_{EM}\alpha_Z  e_Q^2 a_Q^2}{M^2 m_Z^2} (1-R_Z^2)|R_P'(0)|^2, \\
\Gamma(^3P_0 \to Z\gamma) &  = &\frac{32\N \alpha_{EM}\alpha_Z  e_Q^2 v_Q^2}{M^4 (1-R_Z)} (3-R_Z)^2|R_P'(0)|^2, \\
\Gamma(^3P_1 \to Z\gamma) & = & \frac{64\N \alpha_{EM}\alpha_Z  e_Q^2 v_Q^2}{M^2 m_Z^2 (1-R_Z)} (1+R_Z) R_Z^2|R_P'(0)|^2, \\
\Gamma(^3P_2 \to Z\gamma) & = &  \frac{64\N \alpha_{EM}\alpha_Z  e_Q^2 v_Q^2}{5M^2 m_Z^2 (1-R_Z)} (R_Z^2 +3R_Z +6) |R_P'(0)|^2, 
\end{eqnarray}\label{Zgamma}
where the definitions of various quantities can be found in the 
paragraph below Eq.~(\ref{eq:ffbar}).


\subsection{$W^+W^-$}
\begin{eqnarray} 
\Gamma(^1S_0 \to W^+W^-) & = & \frac{\N \alpha_W^2 \beta_W^3}{8M^2(1-2R_W)^2} |R_S(0)|^2, \\
\Gamma(^3S_1 \to W^+W^-) & = & \frac{\N M^2\alpha_W^2 \beta_W^3}{48m_W^4} \bigg\{ \frac{R_W(2-R_W)}{(1-2R_W)^2} - \frac{4R_W(5+6R_W)}{1-2R_W}\bigg(e_Q s_W^2 + \frac{v_Q}{1-R_Z}\bigg)  \nonumber \\
&& + 4(1+20R_W+12R_W^2)\bigg(e_Q s_W^2 + \frac{v_Q}{1-R_Z}\bigg)^2\bigg\}|R_S(0)|^2, \\
\Gamma(^1P_1 \to W^+W^-) & = & \frac{3\N\alpha_W^2 \beta_W}{8M^2m_W^2(1-2R_W)^2}\bigg\{1+\beta_W^2 +2R_W\bigg( 1+ \frac{\beta_W^2}{1-2R_W}\bigg)^2\bigg\} |R_P'(0)|^2, \\
\label{3P0WW}
\Gamma(^3P_0 \to W^+W^-) & = & \frac{\N \alpha_W^2 \beta_W}{4m_W^4} \Bigg\{\Bigg[\frac{1}{1-2R_W} \Bigg( 1- 3R_W + \frac{\beta_W^2 R_W}{1-2R_W}\Bigg) - \frac{3}{1-R_H}(\frac{1}{2} -R_W)\Bigg]^2 \nonumber \\
& & + 2 R_W^2 \Bigg[ \frac{1}{1-2R_W}\Bigg(1-\frac{\beta_W^2}{1-2R_W}\Bigg) - \frac{3}{1-R_H}\Bigg]^2\Bigg\} |R_P'(0)|^2, \\ 
\Gamma(^3P_1 \to W^+W^-) & = & \frac{\N \alpha_W^2 \beta_W^3}{32 m_W^2}\bigg\{ [32 R_W^2+ (3-\beta_W^2)]^2 \bigg( \frac{1}{1-2R_W}- \frac{1}{1-R_Z}\bigg)^2\nonumber \\
&& +4R_W \bigg[ \bigg(\frac{3-4R_W}{(1-2R_W)^2}-\frac{4}{1-R_Z}\bigg)^2 + \frac{\beta_W^4}{(1-2R_W)^4}\bigg]\bigg\} |R_P'(0)|^2,\\ 
\Gamma(^3P_2 \to W^+W^-) & = & \frac{\N \alpha_W^2 \beta_W}{40 m_W^4 (1-2R_W)^2} \bigg\{ \bigg( 1- \frac{2R_W\beta_W^2}{1-2R_W}\bigg)^2 \nonumber \\
 &&+ 6R_W \Bigg[1- \frac{2R_W \beta_W^4}{(1-2R_W)^2} +\bigg( 1- \frac{\beta_W^2}{1-2R_W}\bigg)^2\Bigg] \nonumber \\
 && + 8R_W^2\Bigg[ 6 + \bigg( 1- \frac{\beta_W^2}{1-2R_W}\bigg)^2\Bigg] \bigg\}|R_P'(0)|^2, 
\end{eqnarray}
where $\alpha_W = \alpha_{EM} / s_W^2, \beta_W = \sqrt{1-4R_W}$ 
is the relative velocity of the two $W$'s, $e_Q, v_Q,$ and $a_Q$ are 
the electric charge, vector and axial-vector couplings to the 
$Z$ of the quirk, respectively, and $R_W = m_W^2/M^2$. 
\subsection{$ZZ$}
\begin{eqnarray}
\Gamma(^1S_0 \to ZZ) & = & \frac{4\N(a_Q^2 + v_Q^2)\alpha_Z^2\beta_Z^3}{M^2(1-2R_Z)^2}|R_S(0)|^2, \\
\Gamma(^3S_1 \to ZZ) & = & \frac{8\N a_Q^2 v_Q^2\alpha_Z^2 \beta_Z^5}{3m_Z^2 (1-2R_Z)^2}|R_S(0)|^2, \\
\Gamma(^1P_1 \to ZZ) & = & \frac{32\N a_Q^2 v_Q^2 \alpha_Z^2 \beta_Z^3}{M^2 m_Z^2 (1-2R_Z)^2} |R_P'(0)|^2, \\
\Gamma(^3P_0 \to ZZ) & = & \frac{\N \alpha_Z^2 \beta_Z}{32 m_Z^4} \bigg\{ \bigg( 32a_Q^2 -\frac{3-6R_Z}{1-R_H}-\frac{64R_Z^2v_Q^2}{(1-2R_Z)^2}\bigg)^2 \nonumber \\
&&+ 8R_Z^2 \bigg( \frac{3}{1-R_H} - \frac{32 R_Z v_Q^2}{(1-2R_Z)^2} - \frac{8(3-4R_Z)(a_Q^2-v_Q^2)}{(1-2R_Z)^2}\bigg)^2\bigg\} |R_P'(0)|^2, \\
\Gamma(^3P_1 \to ZZ) & = & \frac{16\N\alpha_Z^2\beta_Z^5}{M^2 m_Z^2(1-2R_Z)^2} \bigg( \frac{2R_Z v_Q^2}{1-2R_Z} - a_Q^2\bigg)^2 |R_P'(0)|^2, \\
\Gamma(^3P_2 \to ZZ) & = & \frac{16\N \alpha_Z^2 \beta_Z}{5m_Z^4} \bigg\{ \bigg( a_Q^2 + v_Q^2\frac{4R_Z^2}{(1-2R_Z)^2}\bigg)^2 + \frac{3R_Z}{(1-2R_Z)^2} \bigg(a_Q^2 + v_Q^2 \frac{2R_Z}{1-2R_Z}\bigg)^2 \nonumber \\
&&+(v_Q^2+a_Q^2)^2 \frac{4R_Z^2}{(1-2R_Z)^2}\bigg( 3 + \frac{2R_Z^2}{(1-2R_Z)^2}\bigg)\bigg\} |R_P'(0)|^2, 
\end{eqnarray}\label{ZZ}
where $\beta_Z = \sqrt{1-4R_Z}$ is the relative velocity between the $Z$'s. 
The definitions of other quantities can be found below 
Eq.~(\ref{eq:ffbar}).


\subsection{$ZH$}
\begin{eqnarray} 
\Gamma(^1S_0 \to ZH ) & = & \frac{\N \alpha_Z^2 a_Q^2 M^2 \beta_{ZH}^3}{4m_Z^4} |R_S(0)|^2, \\
\Gamma(^3S_1 \to ZH) &  = & \frac{\N \alpha_Z^2 v_Q^2 \beta_{ZH}}{6m_Z^2} \bigg\{\bigg( \frac{1-R_H+R_Z}{1-R_H-R_Z} - \frac{2R_Z}{1-R_Z} \bigg)^2 +\frac{R_Z}{2} \bigg( \frac{1-R_H+R_Z}{1-R_Z} - \frac{2}{1-R_H-R_Z}\bigg)^2\bigg\}|R_S(0)|^2, \nonumber \\
&& \\
\Gamma(^1P_1 \to ZH) &  = & \frac{2 \N v_Q^2 \alpha_Z^2 \beta_{ZH}^3}{M^2m_Z^2 (1-R_H-R_Z)^2}|R_P'(0)|^2, \\
\Gamma(^3P_0 \to ZH) &  = &0, \\
\Gamma(^3P_1 \to ZH) & = &\frac{2 \N a_Q^2 \alpha_Z^2 \beta_{ZH}}{m_Z^4} \bigg\{(1-R_H+R_Z)^2\bigg(\frac{R_Z}{1-R_Z} - \frac{1}{1-R_H-R_Z}\bigg)^2 \nonumber \\ 
&& + 8R_Z \bigg( \frac{R_Z}{1-R_Z} - \frac{1}{1-R_H-R_Z} - \frac{\beta_{ZH}^2}{4(1-R_H-R_Z)^2}\bigg)^2 \bigg\} |R_P'(0)|^2, \\
\Gamma(^3P_2 \to ZH) & = & \frac{3\N a_Q^2 \alpha_Z^2 \beta_{ZH}^5}{5M^2 m_Z^2(1-R_H-R_Z)^4}  |R_P'(0)|^2, 
\end{eqnarray}


\subsection{$\gamma H$}
\begin{eqnarray} 
\Gamma(^3S_1 \to \gamma H ) &  = & \frac{\N e_Q^2 \alpha_{EM} \alpha_Z (1-R_H)}{6m_Z^2}|R_S(0)|^2, \\
\Gamma(^1P_1 \to \gamma H ) &  = & \frac{2\N eQ^2\alpha_{EM}\alpha_Z (1-R_H) }{M^2 m_Z^2}|R_P'(0)|^2, \\
\end{eqnarray}

\subsection{$HH$}
\begin{eqnarray} 
\Gamma(^3P_0 \to HH) &  = & \frac{\N \alpha_Z^2 \beta_H}{32 m_Z^4}\bigg(\frac{9 R_H}{1-R_H} -\frac{6}{1-2R_H} + \frac{\beta_H^2}{(1-2R_H)^2}\bigg)^2 |R_P'(0)|^2, \\
\Gamma(^3P_2 \to HH) & =  & \frac{\N \alpha_Z^2\beta_H^5}{80 m_Z^4 (1-2R_H)^4} |R_P'(0)|^2.
\end{eqnarray}

\end{widetext}


\end{document}